\documentclass[fleqn,usenatbib]{mnras}

\usepackage{newtxtext,newtxmath}

\usepackage[T1]{fontenc}

\DeclareRobustCommand{\VAN}[3]{#2}
\let\VANthebibliography\thebibliography
\def\thebibliography{\DeclareRobustCommand{\VAN}[3]{##3}\VANthebibliography}

\usepackage[dvipsnames]{xcolor}
\usepackage{graphicx}	
\usepackage{amsmath}	
\usepackage{float}
\usepackage{subcaption}
\usepackage{booktabs}
\usepackage{verbatim}
\usepackage{caption}
\usepackage{subcaption}
\usepackage{xltabular} 
\usepackage{ragged2e} 
\usepackage{array}  
\usepackage[normalem]{ulem}
\usepackage{hyperref}
\usepackage[english]{babel}
\defcitealias{nava-moreno+24}{NM24}
\renewcommand{\theenumi}{\roman{enumi}}

\newcommand{\TolTEC}{TolTEC}

\title[Bright infrared galaxies in cluster evolution]{Tracing luminous infrared galaxy populations through cluster evolution in a cosmological mock redshift survey}

\author[N. A. Nava-Moreno]{    
    Norma Araceli Nava-Moreno,$^{1}$\thanks{E-mail: aracelinavam@gmail.com}
    Alfredo Montaña,$^{1}$
    Itziar Aretxaga,$^{2}$
    Aldo Rodríguez-Puebla,$^{3}$
    \newauthor
    Vladimir Avila-Reese$^{3}$
    \\
    $^{1}$Instituto Nacional de Astrofísica, Óptica y Electrónica (INAOE), Luis Enrique Erro 1, Sta. Ma. Tonantzintla, 72840, Puebla, Mexico\\
    $^{2}$ Centro de Astrobiologia (CAB), CSIC-INTA, Camino Bajo del Castillo s/n, 28692, Villanueva de la Cañada, Madrid, Spain\\
    $^{3}$Universidad Nacional Aut\'onoma de M\'exico, Instituto de Astronom\'ia, A. P. 70-264, 04510, Ciudad de M\'exico, Mexico}

\date{Accepted XXX. Received YYY; in original form ZZZ}

\pubyear{\the\year{}}

\begin{document}
\label{firstpage}
\pagerange{\pageref{firstpage}--\pageref{lastpage}}
\maketitle

\begin{abstract}
We present GARDENS-Wide, a new 100 square degree mock redshift survey of the dusty star-forming galaxy population based on the MultiDark-\textit{Planck} 2 dark-matter halo simulation. The mock reproduces the observed multiwavelength number counts at 500 $\mu$m, 1.1, 1.4 and 2.0 mm. The large simulated area allows us to identify gravitationally bound systems, trace their assembly histories, and quantify the redshift evolution of their galaxy content and structural extent. We find a strong evolution in the median fractional contribution of galaxy populations within cluster progenitors. Star-forming galaxies account for $\sim$35 per cent of members at low redshift, rising to 60--65 per cent at $z\sim2$ and declining to $\sim$20 per cent by $z\sim5$. LIRGs contribute $\sim$20--40 per cent, peaking near $z\sim2$, whereas ULIRGs remain subdominant ($\lesssim$10 per cent) and HyLIRGs are rare ($<0.5$ per cent). We measure the proto-cluster radius, defined as the maximum 3D comoving distance from the central halo to its most distant member halo. We find that proto-clusters undergo significant contraction over cosmic time: those evolving into rich clusters contract from $\gtrsim$20 comoving Mpc at $z\gtrsim5$ to $\sim$5 comoving Mpc by $z\sim0$, while those evolving into poor clusters evolve from $\sim$13 to $\sim$2 comoving Mpc. The ULIRG population becomes increasingly centrally concentrated at $z>1.5$ for rich proto-clusters, typically confined within the inner 10--65 per cent of the proto-cluster radius. Their projected angular extent is $\sim9$ arcmin at $z\sim5.5$ and $\sim16$ arcmin at $z\sim0$, whereas poor proto-clusters remain more compact ($\sim4$--5 arcmin) across all redshifts. We also provide observational predictions for the TolTEC Large-Scale Structure survey.

\end{abstract}




\section{Introduction}
 
Galaxy clusters are massive ($M\gtrsim 10^{14}\,\rm M_{\sun}$), rare, gravitationally bound systems and represent the most massive virialized structures in the present day Universe \citep{allen+11, Kravtsov+12}. They reside in the most massive dark matter halos and contain from a few tens of member galaxies \citep{Abell1958, Soucail+2015} to several hundreds or even thousands in the most massive nearby systems, together with a diffuse stellar component, and hot intracluster gas \citep{Kravtsov+12}. In the local Universe, clusters are typically observed at low redshift ($z \lesssim 0.5$) as dynamically evolved systems where most galaxies show low star formation activity \citep{dressler1980, Kravtsov+12}.

In the framework of hierarchical structure formation, galaxy clusters assemble over cosmic time through the accretion and merging of smaller halos \citep{white&rees1978,Lacey_Cole1993, Kravtsov+12}. Observational surveys at optical, X-ray, and millimeter wavelengths have extended cluster detections to higher redshifts, confirming the existence of massive clusters out to $z \sim 2$ \citep[e.g.,][]{Wang+2016}. At earlier cosmic times, however, these systems had not yet reached dynamical equilibrium. Their progenitors are observed as proto-clusters, which are overdense regions that will collapse and evolve into present day galaxy clusters \citep{overzier_2016, chiang+2013}.

Proto-clusters trace the early stages of cluster assembly and provide key insight into how massive structures form. They have been identified up to redshifts $z \sim 6$--8 \citep{toshikawa+2012, laporte+22, hashimoto+2023,Witten+2026} and often show enhanced star formation activity compared to the field \citep{chiang+2017, overzier_2016}. Studying proto-clusters is therefore essential to understand how galaxy clusters assemble their mass and how galaxy populations evolve in dense environments. However, identifying proto-clusters observationally remains challenging. Unlike virialized clusters, proto-clusters are extended structures that can span tens of comoving megaparsecs and have not yet collapsed into compact systems \citep{chiang+2013, overzier_2016}. As a result, their galaxy overdensities can be modest and can be difficult to distinguish from the field
 \citep[e.g.][]{chiang+2013, muldrew+15, overzier_2016}. In addition, their large spatial extent requires wide-area surveys to identify their full structure. Accurate redshift measurements are also needed to separate proto-cluster members from foreground and background galaxies. Furthermore, a significant fraction of star formation in these systems is obscured by dust, making many proto-cluster galaxies difficult to detect at optical wavelengths \citep{overzier_2016, Casey_2016}. Dusty star-forming galaxies (DSFGs) may therefore provide an efficient way to trace proto-cluster environments, as suggested by several DSFG-rich overdensities identified at high redshift \citep[e.g.,][]{casey+2015, kato+2016, oteo+2018, araya-araya+2024}. As a result, observations at far-infrared and submillimetre wavelengths are essential to obtain a more complete census of star-forming activity in these environments. However, such observations face their own challenges, including limited angular resolution at high redshift, source confusion, and the difficulty of combining both wide area and sufficient depth, which further complicate the detection and characterization of proto-clusters.

In this context, cosmological simulations provide a crucial framework to overcome these observational limitations. By tracing the hierarchical growth of dark matter halos across cosmic time, simulations allow us to identify the progenitors of present-day massive clusters and to follow their evolution in full three-dimensional space \citep[e.g.,][]{chiang+2013, chiang+2017, muldrew+15, overzier_2016, Remus+2023, Baxter+2025}. They enable a direct connection between high-redshift overdensities and their $z=0$ descendants, while providing access to physical properties, merger histories and environments, which are difficult to measure observationally.

In this work, we present GARDENS-Wide, the wide-area release of the GARDENS (Galaxies Across Redshift with Dust Emission from Numerical Simulations)\footnote{GARDENS-Wide is publicly available for the community to facilitate comparisons and further studies.} catalogues, a cosmological mock redshift survey of the DSFG population designed to reproduce its infrared emission and large-scale distribution over 100 square degrees. The catalogue is based on a dark matter halo simulation within a cosmological $N$-body framework and includes the effects of gravitational lensing. It builds upon a previous higher-resolution, smaller-area realization (5.3 square degrees) presented in \citet[][hereafter \citetalias{nava-moreno+24}]{nava-moreno+24}, which we designate as GARDENS-Deep. The realization introduced in this paper corresponds to the wider-area implementation of the model. This larger catalogue enables the identification and statistical study of galaxy clusters and their progenitors.

The larger simulated area of GARDENS-Wide allows us to identify gravitationally bound systems, trace the assembly histories, and study the galaxy populations in proto-clusters. We measure their three-dimensional comoving radial extents and their redshift dependent angular sizes for the star-forming galaxy population, with particular emphasis on luminous infrared systems, and use the resulting catalogue to make predictions for forthcoming TolTEC\footnote{http://toltec.astro.umass.edu/} \citep{wilson+2020} observations. TolTEC is a submillimetre wavelength camera installed on the 50 m Large Millimeter Telescope \citep[LMT\footnote{http://lmtgtm.org/};][]{hughes+2020}, observing simultaneously at 1.1, 1.4 and 2.0 mm with angular resolutions of 5.0, 6.3 and 9.5 arcsec, respectively. TolTEC will carry out two major extragalactic legacy surveys: a deep, small area Ultra-Deep Survey (UDS) and a wide area Large-Scale Structure Survey \citep[LSS;][]{montana+2019}. The UDS will reach $1\sigma$ depths of 0.025, 0.018 and 0.012 mJy beam$^{-1}$ over $\sim$0.8 square degree, enabling the detection of typical star-forming galaxies with star formation rates (SFRs) $\gtrsim$10 M$_{\sun}$~yr$^{-1}$. In contrast, the LSS survey will cover a total area of $\sim$40–60 square degree, reaching $1\sigma$ depths of 0.25, 0.18, and 0.12 mJy beam$^{-1}$, enabling the detection of luminous infrared galaxies with total infrared luminosities ($L_{\rm IR}$) $\gtrsim$ $10^{12}$ L$_{\sun}$ and SFRs $\gtrsim$ 100 M$_{\sun}$~yr$^{-1}$. The LSS will probe the large-scale distribution of DSFGs and their connection to the cosmic web \citep{montana+2019}. In this work, we focus on providing predictions relevant for the LSS survey.

This paper is organized as follows. In Section \ref{sec:mock}, we describe the construction of the mock catalogue. In Section \ref{sec:clusters_history}, we identify gravitationally bound systems as pairs, groups, and clusters within our lightcone. We then trace the assembly histories of groups and clusters toward higher redshift and quantify the contribution of proto-clusters to the cosmic star formation rate density. In Section \ref{sec:cluster_analysis}, we characterize the galaxy population and spatial properties of proto-clusters and groups, including  a brief characterization of galaxy pairs. In Section \ref{sec:toltec_predictions}, we present predictions for forthcoming TolTEC observations. Finally, in Section \ref{sec:summary_conclusions}, we summarize our main results and conclusions. The total $L_{\rm IR}$ is defined as the luminosity integrated over the rest-frame 8–1000 $\mu$m wavelength range. Throughout the paper, we adopt a \citet{chabrier2003} IMF, as well as a flat $\Lambda$CDM cosmology with: $\Omega_{m}$ = 0.307, $\Omega_{\Lambda}$ = 0.693 and $h$ = 0.678 \citep{planck+2016}.

\section{The mock redshift survey} \label{sec:mock}

In this section, we summarize the construction of our 100 square degree mock redshift survey, which follows the methodology introduced in \citetalias{nava-moreno+24}. For completeness, we provide a brief overview of the main steps and highlight updates implemented here.

\subsection{Lightcone}

To build our lightcone we follow the methodology in \citetalias{nava-moreno+24}, using the MultiDark-\textit{Planck} 2 cosmological N-body simulation \citep[MDPL2,][see also, \citealp{RP+2016b} for a comparison with other simulations]{klypin+2016} with periodic boundary conditions. MDPL2 has a side length of 1 $h^{-1}$Gpc, and contains 3840$^{3}$ dark matter particles of 1.51$\times 10^{9}$ $h^{-1}$M$_{{\sun}}$. We replicate the box simulation 7 times according to equation (1) in \citetalias{nava-moreno+24} to span a redshift range from 0 to 10, yielding a lightcone that covers an area of 100 square degree and a volume $V\approx2.3\ h^{-3}$Gpc$^{3}$.

Since our goal is to study DSFGs, we limit the entire analysis in this paper to $0 \le z \le 7$, as the presence and abundance of DSFGs at very high redshift remain uncertain.

To populate dark matter halos with galaxies, we adopted an updated version of the semi-empirical modelling of the galaxy–halo connection developed by \citet{RP+2016a, RP+2017}, in which each (sub)halo\footnote{Here, we will use the term '(sub)halo' to refer to both haloes and subhaloes.} in the simulation is assigned a single galaxy. This implementation follows a generalized \citep{RP+2012,RP+2013} Subhalo Abundance Matching (SHAM) approach, where the cumulative number densities of (sub)halos and galaxies are matched through a monotonic relation between a halo property, such as the maximum circular velocity along the main progenitor branch ($V_{\rm peak}$), and a galaxy property, namely the stellar mass ($M_{\star}$). The SFRs are inferred from the growth histories of the host halos, decomposing the stellar mass growth into in-situ star formation and ex-situ contributions from galaxy mergers. Finally, the growth history of each galaxy in the MPL2 simulation was derived to reproduce the evolution of the galaxy stellar mass function between $0\lesssim z\lesssim10$, the fraction of quiescent galaxies between $0\lesssim z\lesssim4$, the mean of the star-forming main sequence of galaxies from $0\lesssim z \lesssim10$ and the UV and IR luminosity functions since $0\lesssim z \lesssim11$ and $0\lesssim z \lesssim4.5$, respectively (for details, see Rodríguez-Puebla et al., in prep.; see also \citealt{RP24}).

\subsection{Star-Forming Galaxies and Their Infrared Properties}\label{subsec:2.2}

Galaxies in our mock catalogue are classified as star-forming or quiescent by computing their specific star formation rate ($\mathrm{sSFR}$ = $\mathrm{SFR} / \textrm{M}_{\star}$) and applying the redshift-dependent $\mathrm{sSFR}$ threshold of \citet{pacifici+2016}, which parametrizes the evolution of the boundary between star-forming and quiescent galaxies as a function of redshift. The main sequence of star-forming galaxies is derived through an iterative sigma-clipping procedure in stellar-mass and redshift bins, ensuring a robust separation between the two populations.

The infrared properties of galaxies, including the dust–obscured star formation rate ($\mathrm{SFR}_{\rm IR}$), $L_{\rm IR}$, the dust temperature ($T_{\rm dust}$), and the observed flux densities, are derived using theoretical and empirical relations from the literature. Here, $\mathrm{SFR}_{\rm IR}$ corresponds to the component of star formation traced by the total infrared luminosity, $L_{\rm IR}$ (8–1000 $\mu$m). To obtain this parameter, we model the obscured fraction ($f_{\rm obs} = \mathrm{SFR}_{\mathrm{IR}}\text{/}\mathrm{SFR}$) as a function of stellar mass and redshift by combining different observational and theoretical constraints. We adopt the empirical relation of \citet{whitaker+2017}, use the measurements of the total and infrared star formation rate densities ($\mathrm{SFRD}$) from \citet{dunlop+2017} to estimate $f_{\rm obs}$, and apply the theoretical prescription of \citet{R-P+2020} for high-mass galaxies at low redshift. This allows us to estimate $\mathrm{SFR}_{\rm IR}$ via
$\mathrm{SFR}_{\mathrm{IR}} = f_{\rm obs} \cdot \mathrm{SFR}$,
and then derive infrared luminosities for our galaxies using the inverse of the \citet{kennicutt_1998} relation scaled to a \citet{chabrier2003} IMF: $\mathrm{SFR}_{\mathrm{IR}}\ \text{/}{\rm M_{\sun}yr^{-1}}
= 1.09 \times 10^{-10}\, L_{\mathrm{IR}}\ \text{/}\rm L_{\sun}$.

A few galaxies in our mock catalogue exhibit extremely high total star formation rates, reaching values of up to $\sim20,000$~M$_{\sun}$~yr$^{-1}$. These values are likely artifacts of the catalogue construction, arising from the combination of observational scatter, unusually rapid halo growth in the simulation, or occasional inaccuracies in the merger trees. To avoid these unphysical outliers, we impose, prior to applying lensing effects, an upper limit of 6,000 M${\sun}$~yr$^{-1}$ on the $\mathrm{SFR}_{\rm IR}$ of galaxies, motivated by recent observational results from \citet{quiros-rojas+24}, which show that high redshift submillimetre galaxies do not exceed this value. This choice updates the 2,000~M$_{\sun}$~yr$^{-1}$ threshold adopted in \citetalias{nava-moreno+24} and affects only 23 galaxies in the full 100 square degree mock catalogue.

Dust temperatures are assigned based on the intrinsic $L_{\rm IR}$ of the galaxies using the empirical scaling of \citet{Casey+2018a}, which relates $L_{\rm IR}$ to the rest-frame peak wavelength ($\lambda_{\rm{peak}}$) of the spectral energy distribution (SED). An approximation to Wien’s law is then used to infer the dust temperature. However, the $T_{\rm dust}$–$L_{\rm IR}$ relation exhibits significant intrinsic scatter \citep[e.g.][]{Casey+2012b, Chapman+2005}. To account for this and better represent the diversity of galaxy dust temperatures, we include the dispersion reported by \citet{Casey+2012b}. The resulting temperatures are then corrected for the effects of the heating of the cosmic microwave background, which become significant at high redshifts \citep[$z>4$,][]{dacunha+2013}.

To model the emission of interstellar dust and estimate intrinsic flux densities, we adopt a gray-body SED in which the dust emissivity index $\beta$ is drawn from a Gaussian distribution with mean $\beta = 2.2$ and dispersion $\sigma = 0.34$. This approach replaces the fixed value $\beta = 1.8$ adopted for GARDENS-deep in \citetalias{nava-moreno+24}. The adopted dispersion is derived from the observational constraints reported by \citet{ward+2024} who find a medium value of $\beta=1.96$ for a optically thin model with a confidence interval (16-84 per cent) of 1.67 to 2.35 for the DSFG population. We explored several central values of $\beta$, each combined with the same dispersion, and found that $\beta = 2.2$ provides the best agreement with the observed number counts at 1.1, 1.4, and 2.0~mm. Based on this test, we adopt $\beta = 2.2$ with $\sigma = 0.34$ as our fiducial model, and assign to each galaxy in the mock catalogue a value of $\beta$ randomly drawn from this distribution, allowing us to capture the observed diversity in dust properties among galaxies. Our adopted value is consistent with the typical range $\beta \sim 1.5$–2.5 reported for DSFGs \citep[e.g.][]{daCunha+2021, ward+2024, McKay+2023, bendo+2025}.

Finally, as in \citetalias{nava-moreno+24}, this new mock redshift survey also includes the effects of gravitational lensing, using a simple point mass lensing model. We adopt a minimum amplification of $\mu = 1.2$, taking into account only the effect of strong gravitational lenses, which can significantly impact the number counts.

\subsection{Characterization of the catalogue}

Our final 100 square degree mock catalogue contains only galaxies classified as star-forming, whereas both star-forming and quiescent populations are retained for the analysis of galaxy clusters presented later in this work (Sections \ref{sec:clusters_history} and \ref{sec:cluster_analysis}). Table \ref{tab:comparing_simulations} summarizes the main properties and parameter ranges of this new mock survey, and contrasts them with those of the catalogue presented in \citetalias{nava-moreno+24}.

\begin{table}
\centering
\caption{Comparison of the main properties of the GARDENS-Wide mock catalogue presented in this work with those of GARDENS-Deep (\citetalias{nava-moreno+24}). The reported quantities correspond to intrinsic (unlensed) values, while the lensing amplification is given by the parameter $\mu$.}
\label{tab:comparing_simulations}
\begin{tabular}{lcccc}
\toprule
  & \multicolumn{2}{c}{GARDENS-Wide} & \multicolumn{2}{c}{GARDENS-Deep} \\
Parameter & \multicolumn{2}{c}{This work (100 deg$^2$)} & \multicolumn{2}{c}{NM24 (5.3 deg$^2$)} \\
\cmidrule(lr){2-3} \cmidrule(lr){4-5}
          & Min. & Max. & Min. & Max. \\
\midrule
Box simulation & \multicolumn{2}{c}{MultiDark-Planck 2} & \multicolumn{2}{c}{Bolshoi-Planck} \\
Redshift ($z$) & 0.002 & 7.00 & 0.02 & 7.00 \\
log ($M_{\rm vir}$/M$_{\sun}$) & 10.38 & 14.84 & 9.70 & 13.89 \\
log ($M_\star$/M$_{\sun}$) & 9.50 & 12.04 & 8.75 & 11.83 \\
Total SFR [M$_{\sun}$\,yr$^{-1}$] & 0.46 & 22,132 & 0.077 & 7,848 \\
SFR$_{\rm IR}$ [M$_{\sun}$\,yr$^{-1}$] & 0.26 & 6,000 & 0.015 & 2,000 \\
log ($L_{\rm IR}$/L$_{\sun}$) & 9.38 & 14.23 & 8.15 & 13.26 \\
$T_{\rm dust}$ [K] & 19.0 & 100.6 & 20.8 & 56.4 \\
Magnification ($\mu$) & 1.2 & 175 & 1.2 & 272 \\
\bottomrule
\end{tabular}
\end{table}

\begin{figure*}
  \centering
  \begin{subfigure}[b]{0.48\textwidth}
    \includegraphics[width=\linewidth]{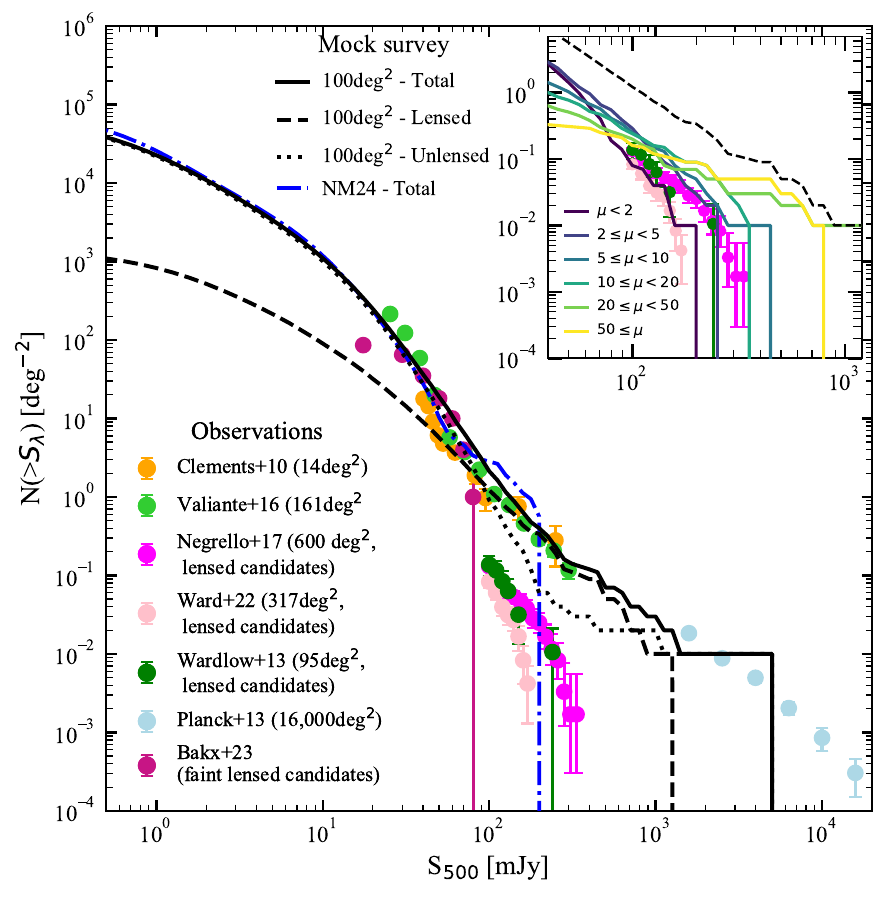}
  \end{subfigure}
  \hfill
  \begin{subfigure}[b]{0.48\textwidth}
    \includegraphics[width=\linewidth]{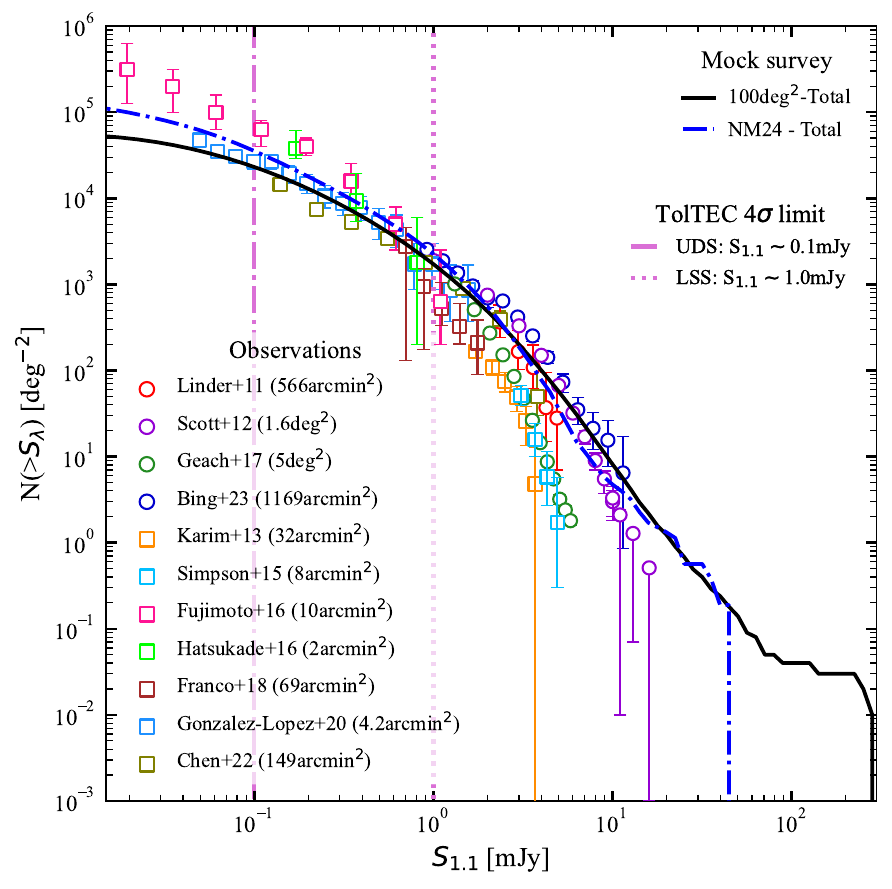}
  \end{subfigure}
  
  \vspace{0.3em} 

  \begin{subfigure}[b]{0.48\textwidth}
    \includegraphics[width=\linewidth]{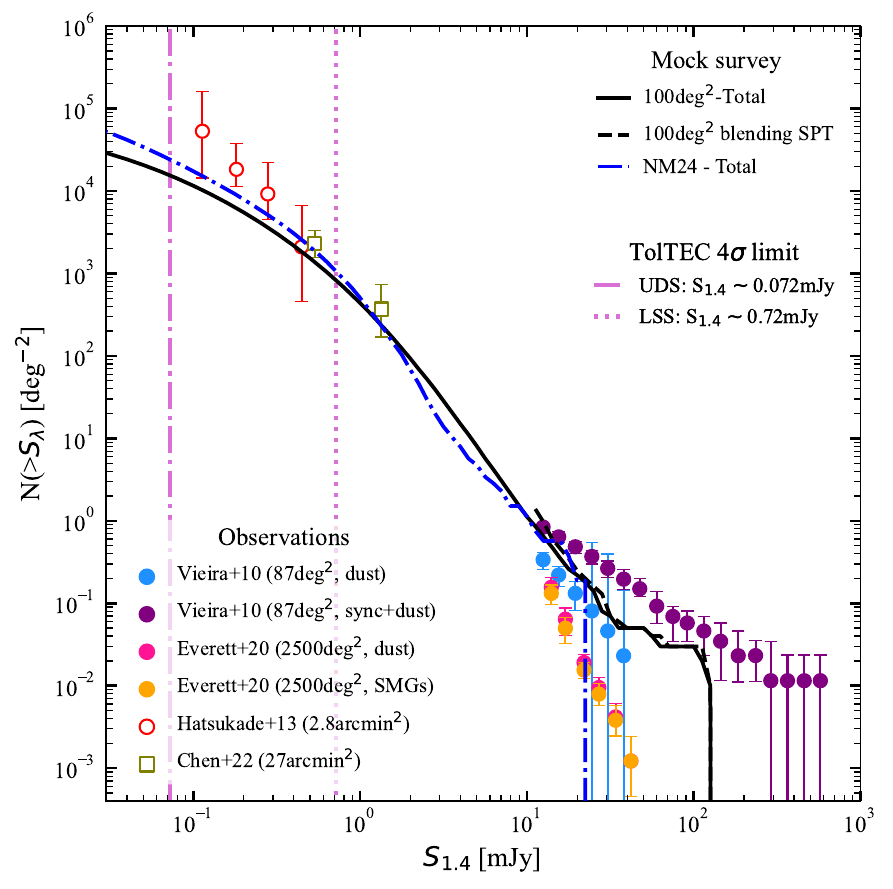}
  \end{subfigure}
  \hfill
  \begin{subfigure}[b]{0.48\textwidth}
    \includegraphics[width=\linewidth]{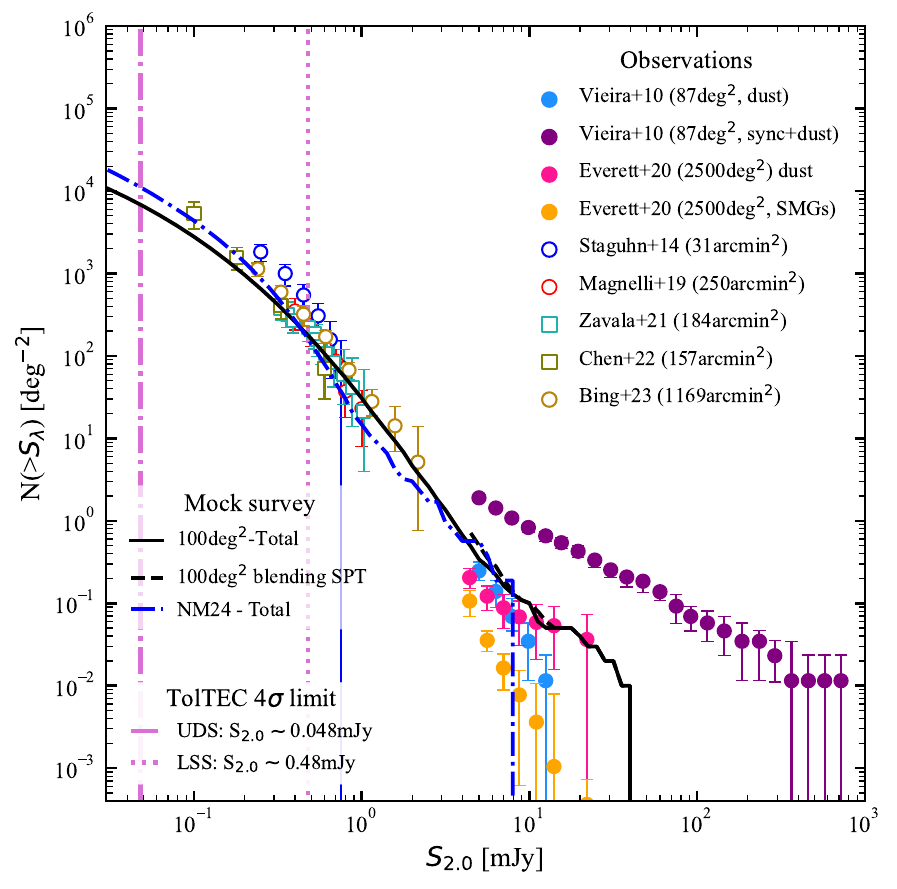}
  \end{subfigure}
  \caption{Cumulative number counts at 500 $\mu$m, 1.1, 1.4 and 2.0 mm measured within our simulated area of 100 square degrees.
\textbf{Upper left:} Total number counts at 500 $\mu$m (black solid line), separated into lensed ($\mu>1.2$, black dashed line) and unlensed (black dotted line) galaxies. These are compared with several large-area surveys at 500 $\mu$m \citep{clements+2010, planck+2013, wardlow+2013, valiante+2016, negrello+2017, ward+2022, bakx+2023}. The inset panel shows the contribution of mock lensed sources to the number counts at different magnifications. \textbf{Upper right:} Number counts at 1.1 mm (black solid line) compared with interferometric (squares) and single-dish (circles) observations \citep{Lindner+2011, scott+2012, geach+2017, bing+2023, karim+2013, simpson+2015, fujimoto+2016, hatsukade+2016, franco+2018, gonzalez-lopez+2020, chen+2022}.
\textbf{Lower left:} Number counts at 1.4 mm (black solid line) compared with our mock catalogue blended at the SPT beam resolution (black dashed line). Observational results affected by gravitational lensing are shown as blue, pink, and orange dots for dusty sources, and purple dots for mixed dusty + synchrotron populations \citep{vieira+2010, everett+2020}. Counts of galaxies not expected to be dominated by a strongly lensed population are shown as circles nad empty squares \citep{hatsukade+2013, chen+2022}. \textbf{Lower right:} Number counts at 2.0 mm (black solid line), also compared with our mock catalogue convolved with the SPT beam (black dashed line) and with observations from \citet{vieira+2010, everett+2020}, shown as in the previous panel. Circles and empty squares denote counts of galaxies not expected to be dominated by a strongly lensed population \citep{staguhn+2014, magnelli+2019, zavala+2021, chen+2022, bing+2023}. Vertical violet lines mark the $4\sigma$ detection limits at 1.1, 1.4, and 2.0 mm for the planned UDS and LSS TolTEC surveys.}
  \label{fig:ncounts}
\end{figure*}

Figure \ref{fig:ncounts} presents the integrated number counts at 500 $\mu$m, 1.1, 1.4, and 2.0~mm measured in our 100~square degree mock catalogue. Overall, the counts are consistent with the counts presented in \citetalias{nava-moreno+24}. The divergence at lower flux densities reflects the higher halo mass resolution in \citetalias{nava-moreno+24}, which reproduces a larger number of faint galaxies, whereas this work is better suited for sampling the bright end of the counts. Given that the present mock sample covers a wider area, in this paper we focus our comparison primarily to large-area surveys, where gravitational lensing amplification can play an important role.

We compare our number counts with a broad compilation of observational results. At 500~$\mu$m, our mock catalogue is compared with measurements from wide-area \textit{Herschel} surveys, including H-ATLAS \citep{clements+2010, valiante+2016, negrello+2017, ward+2022}, HerMES \citep{wardlow+2013}, and \textit{Planck} \citep{planck+2013}, as well as with ALMA follow-up observations of lensed \textit{Herschel} sources \citep{bakx+2023}.

Our number counts at 500 $\mu$m reproduce the observed total counts over a wide range of flux densities. The agreement with large area surveys such as H-ATLAS and the all-sky \textit{Planck} survey indicates that the adopted modelling of infrared properties produces a realistic population of dusty star-forming galaxies. At the bright end ($S_{500}\gtrsim100$ mJy), where the contribution from gravitationally lensed systems becomes significant, our model reproduces the observed slope of the number counts, although it slightly overpredicts the abundance of lensed sources compared to surveys specifically targeting such systems. Observational studies find typical magnifications in the range $\mu\sim2$–16 \citep{negrello+2017} and $\mu\sim2$–23 \citep{wardlow+2013}. When we split the model counts by magnification (inset of Figure~\ref{fig:ncounts}), we find that sources with moderate amplifications ($\mu\sim2$–10) are consistent with the observed lensed population, whereas bins with higher magnifications ($\mu\sim10$–20) lie above the observational constraints, indicating an excess of highly magnified systems in our model. This discrepancy likely reflects limitations of the point-mass lensing approximation, which does not fully capture the complexity of realistic lens mass distributions and biases the magnification distribution toward higher values.

At 1.1, 1.4, and 2.0 mm, we compare our results with number counts from both interferometric and single-dish surveys \citep{Lindner+2011, scott+2012, karim+2013, hatsukade+2013, staguhn+2014, simpson+2015, fujimoto+2016, hatsukade+2016, geach+2017, magnelli+2019, gonzalez-lopez+2020, franco+2020, zavala+2021, chen+2022, bing+2023}. At 1.4 and 2.0 mm, we additionally include measurements affected by gravitational lensing of the South Pole Telescope (SPT), which identify a large franction of gravitational lensed systems. \citep{vieira+2010, everett+2020}.

Our number counts reproduce the overall compilation of interferometric and single–dish observational data. At the bright end of the 1.4 and 2.0~mm counts, where dust–dominated sources prevail, our results remain broadly consistent with the observed number counts.

In order to test whether source blending could artificially increase the number counts at high flux densities, we simulated the effect of blending using the beam sizes of the SPT observations at 1.4 and 2.0 mm (FWHM $\sim$1.0 and 1.2 arcmin, respectively). We first selected galaxies above the SPT detection limits at each wavelength. For each of these galaxies, we defined a circular aperture corresponding to the beam and summed the fluxes of all neighbouring galaxies within this apertures, regardless of their flux. This experiment shows no significant impact on the resulting number counts for bright galaxies comparable to those detected by SPT. We note that this test is limited to this flux regime and does not address the impact on fainter sources.

In addition to the number counts, we verified other key statistical properties of the new catalogue. Specifically, we compared the $\mathrm{SFRD}$, the infrared luminosity function, and the redshift distribution with those from \citetalias{nava-moreno+24}. The results are consistent with the previous mock GARDENS-Deep, confirming that the inclusion of dispersions in $T_{\rm dust}$ and $\beta$, as well as the adopted central value of $\beta$, does not significantly affect the global statistical trends. Minor differences arise from the larger area and lower mass resolution of the present catalogue, which extends the bright end of the infrared luminosity function while yielding a slightly lower $\mathrm{SFRD}$ due to the reduced sampling of faint galaxies.

\section{Identification of structures and their histories} \label{sec:clusters_history}

In this section, we analyze gravitationally bound systems in our mock catalogue, from galaxy pairs to massive clusters. We examine their growth histories across cosmic time and explore the properties and evolution of their galaxy populations. We quantify their contribution to the cosmic star formation rate density. The analysis uses the full mock catalogue, including both quiescent and star-forming galaxies.

\subsection{Identification of Bound Structures Within GARDENS-Wide}\label{sec:halo_systems}

The MDPL2 halo and subhalo catalogues were generated using the ROCKSTAR halo finder algorithm \citep{behroozi+2013b}, which identifies gravitationally bound structures based on phase–space densities, followed by the algorithm CONSISTENT TREES \citep{behroozi+2013c} to construct the corresponding merger trees. The last one enforces physically and temporally consistent halo associations across snapshots, ensuring a reliable reconstruction of their evolutionary histories.

In the resulting catalogue, each halo has an assigned ID and UPID. The first is a unique identifier for each halo at a given snapshot in the lightcone. The UPID encodes the hierarchical structure of the catalogue and specifies the ID of the main host halo to which each subhalo is associated. Halos with UPID equal to a given ID are classified as subhalos of that host halo. This UPID–ID relation defines the hierarchical organization of structures at each snapshot and is used in this work to determine system membership. Importantly, this hierarchical association does not imply that all member halos are dynamically virialized within the host halo, as systems may include both virialized and infalling components.

Building on this framework, we identify bound systems across the full redshift range of our mock catalogue using the hierarchical structure of the MDPL2 halo catalogue. Each system is defined by a host halo and all associated subhalos connected through the UPID–ID relation. We then classify these systems according to the number of subhalos (or satellite galaxies) associated with the same central halo.\footnote{Throughout this work, the central halo refers to the main (host) halo of the system, as identified in the merger tree, and should not be interpreted as the geometric centre of the system.} We distinguish five categories: isolated halos (defined as halos with no other gravitationally bound halo identified within our lightcone at the mass resolution of the catalogue), pairs, groups (3–10 members), poor clusters (11–50 members), and rich clusters (more than 50 members). This classification enables us to follow the evolution of structures of different richness across cosmic time and to characterize the luminous infrared galaxy population within them across different environments. Given the low stellar mass limit of our mock catalogue, the above numbers refer to galaxies with masses greater than log($M_\star/M_{\sun}$)=9.5, see Table \ref{tab:comparing_simulations}.

\begin{figure}
  \centering
    \subfloat{\includegraphics[width=\columnwidth]{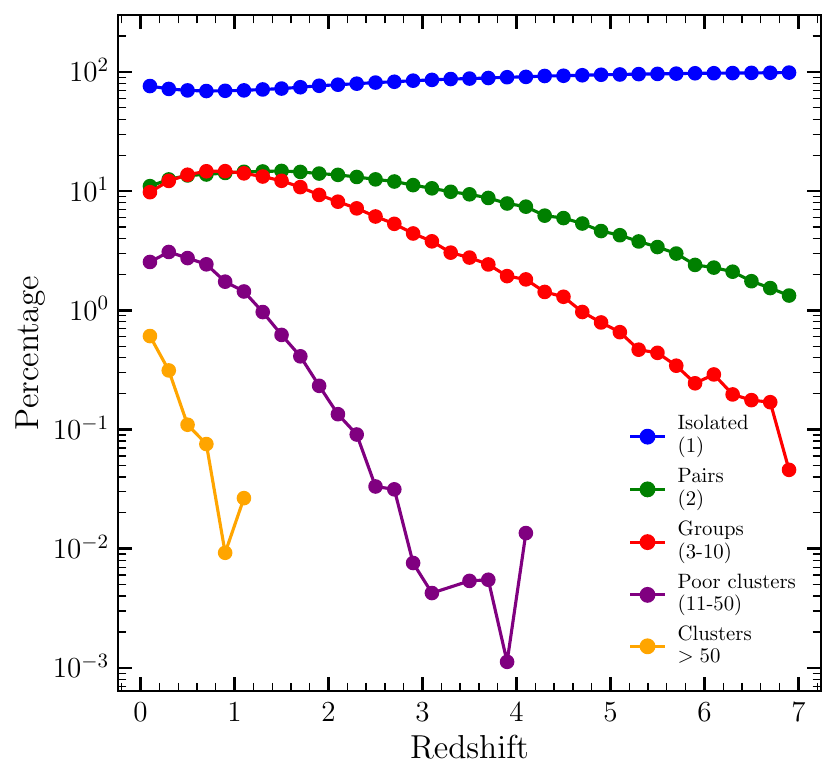}}    
  \caption{Percentage of galaxies in each association category relative to the total number of galaxies within each redshift bin, based on our 100 square degree mock redshift survey. Categories are defined by the number of members: isolated (1), pairs (2), groups (3--10), poor clusters (11--50), and rich clusters ($>50$). The dominant population at all redshifts corresponds to isolated galaxies, followed by pairs, small groups, poor clusters and clusters, consistent with hierarchical structure formation.}
  \label{fig:percent_classification}
\end{figure}

Figure \ref{fig:percent_classification} shows the percentage of galaxies in each category as a function of redshift bin, where the fractions are calculated relative to the total number of galaxies in that bin. As expected, isolated galaxies dominate across all redshifts, increasing from approximately 70 per cent at $z < 1.5$ to 98 per cent at $z > 4$. Paired galaxies are the second most common configuration, peaking at 14 per cent in the range $0.5 < z < 1.0$ and declining to 1.5 per cent by $z = 7$. Galaxies in groups also peak at 14 per cent within $0.5 < z < 1.0$, but their abundance drops to just 0.15 per cent at $z = 7$. Galaxies in poor clusters reach a maximum of 4 per cent between $z \approx 0$ and $z = 0.5$, and become negligible at $z > 2$; beyond $z > 4.5$, poor clusters are no longer present in the mock survey. Galaxies in rich clusters contribute approximately 0.2 per cent at $z < 0.5$, and are absent at $z > 1.1$. 

\begin{figure}
  \centering
    \subfloat{\includegraphics[width=\columnwidth]{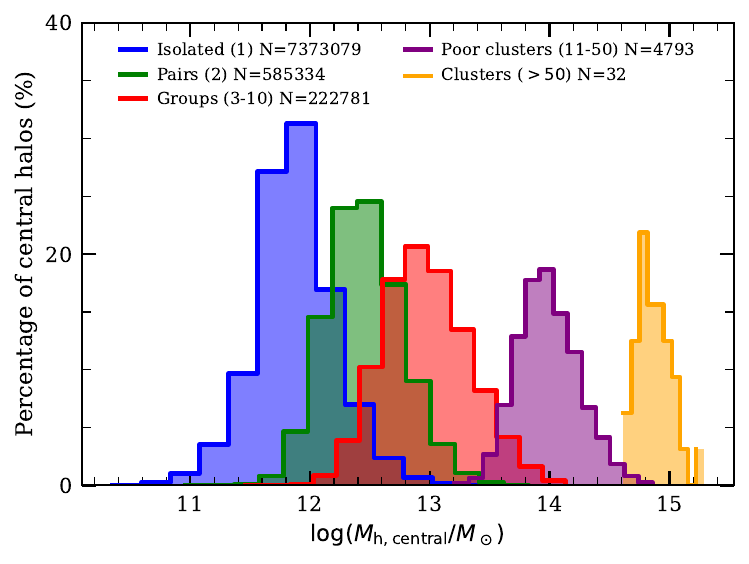}}    
  \caption{Mass distributions of central halos in each category, as identified in our mock survey at all redshifts. The $y$-axis shows the percentage of halos in each mass bin relative to the total number in that category (N). The distributions show that isolated halos dominate the low-mass end, while richer associations are progressively shifted toward higher halo masses, highlighting the hierarchical growth of structures.}
  \label{fig:dist_mass}
\end{figure}

This trend is consistent with expectations from hierarchical structure formation: although the hierarchical growth itself is revealed through the time evolution of halos, at fixed redshift more massive central halos are statistically associated with richer environments. Figure \ref{fig:dist_mass} shows the mass distributions of central halos in each category, with central halos in isolated systems occupying the low-mass end, while central halos in pairs, groups, and clusters are increasingly
shifted towards higher masses. Table \ref{tab:dist_mass_stats} lists the number of associations in each category (e.g. number of pairs, groups, clusters) found within our 100 square degree mock survey, together with the median central halo mass and the interval spanned by each mass distribution.

\begin{table}
\centering
\caption{Mass statistics of central halos for each category in our 100 deg$^2$ mock catalogue. N corresponds to the number of associations found in the full lightcone. The median halo mass ($M_{\rm median}$) and range ($M_{\rm range}$; minimum and maximum halo mass values) are given in log($M_{\rm h,central}$/M$_{\sun}$).}

\label{tab:dist_mass_stats}
\begin{tabular}{lccc}
\toprule
Category & N & $M_{\rm median}$ & $M_{\rm range}$ \\
\midrule
Isolated (1 member) & 7,373,079 & 11.9 & [10.3, 14.0] \\
Pairs (2 members) & 585,334 & 12.4 & [11.0, 14.0] \\
Groups (3--10 members) & 222,781 & 13.0 & [11.5, 14.3] \\
Poor clusters (11--50) & 4,793 & 14.0 & [13.2, 15.0] \\
Rich clusters ($>50$) & 32 & 14.9 & [14.6, 15.3] \\
\bottomrule
\end{tabular}
\end{table}

\subsection{Tracing Assembly Histories}

CONSISTENT TREES enables us to reconstruct the complete merger history of halos and subhalos, and therefore of central and satellite galaxies, by linking progenitors and descendants through unique merger tree (Tree\_root\_ID) and (sub)halo IDs \citep[see Appendix~B of][for further discussion on the tree walking]{RP+2016b}. The Tree\_root\_ID is used to connect halos across different snapshots through the merger tree, identifying all progenitors and descendants of a given halo. This allows us to reconstruct the assembly history of each system by tracing how individual halos grow, merge, and are accreted over cosmic time.

The identification of bound systems in a lightcone is non-trivial, as halos that form a single virialized system at a given epoch may originate from multiple, independent progenitor halos. To consistently track all contributing components across cosmic time, we review the history of each member in the structure.

We first select a bound system at the redshift where it is identified in the lightcone, defined by a central halo and all associated subhalos connected through the UPID–ID relation (see Section \ref{sec:halo_systems}). To reconstruct its formation history, we then trace back all (sub)haloes that belong to the same evolutionary lineage across earlier epochs by walking within their merger trees, capturing the growth of the system through mergers and accretion. While a bound system at a given snapshot contains a single central halo by construction, its past assembly may involve multiple central halos. These correspond to smaller, independent systems that merged over time to form the present-day structure. As a result, not all galaxies currently residing in the system necessarily originate from a single evolutionary path; instead, they may have formed in distinct progenitor haloes that only recently became part of the same bound structure.

To account for this complexity, we iteratively identify all central haloes associated with the system at earlier times and include their corresponding satellite populations, above the stellar mass limit resolution from Table \ref{tab:comparing_simulations}. Whenever a newly identified component follows a different evolutionary track, its full formation history is incorporated. This procedure is repeated until all contributing progenitors are consistently included, ensuring that the full merger history of the system is recovered. We call this set of halos, which trace the assembly history of the clusters at each stage, proto-clusters.

Since these halos host galaxies, our analysis extends to studying the properties of the galaxies within them. In particular, we aim to characterize the population of infrared-bright galaxies using their intrinsic (i.e. unlensed) infrared luminosities, namely Luminous Infrared Galaxies (LIRGs, $10^{11}\;{\rm L_{\sun}}\leq L_{\rm IR}< 10^{12}\; \rm L_{\sun}$), Ultra-Luminous Infrared Galaxies (ULIRGs, $10^{12}\; {\rm L_{\sun}}\leq L_{\rm IR}< 10^{13}\; \rm L_{\sun}$), and Hyper-Luminous Infrared Galaxies (HyLIRGs, $L_{\rm IR}\geq 10^{13}\; \rm L_{\sun}$), which together trace the DSFG population within these associations.

\begin{figure*}
  \centering
  \includegraphics[width=\textwidth]{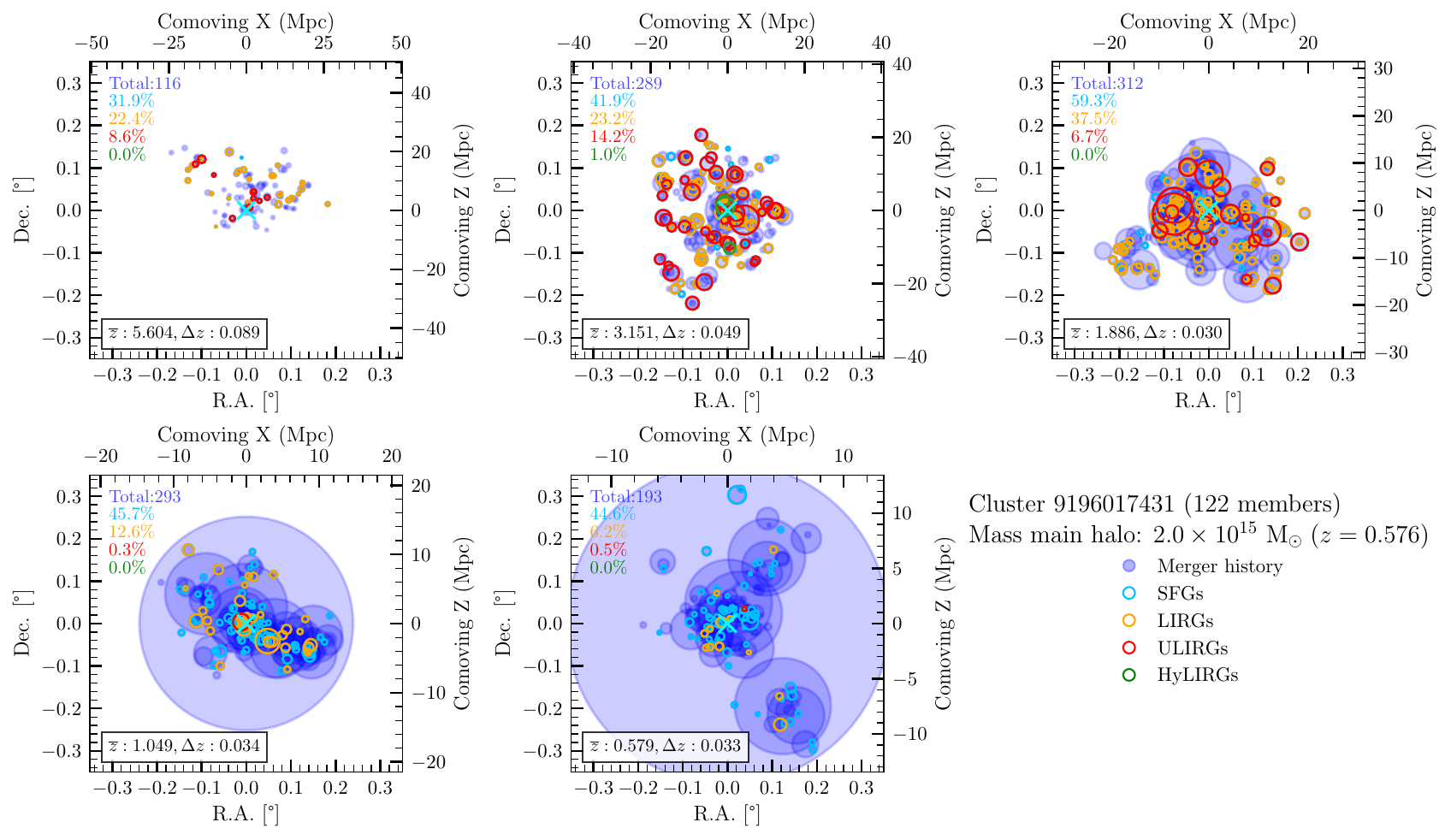}
  \caption{Assembly history of the richest cluster in our $100$ square degree lightcone across several snapshots. At $z=0.576$, the main halo has a mass of $2.0\times10^{15}\,{\rm M}_{\sun}$ and contains 122 cluster members. At this redshift, 71 additional progenitor halos that are not yet cluster members but will eventually assemble into the final cluster at $z=0$, resulting in a total of 193 progenitor halos at $z=0.576$. All halos participating in the assembly at each snapshot are shown as transparent blue filled circles, including star-forming and quiescent galaxies; their size scales with their respective halo mass. All halo positions have been recentered so that the central halo lies at (0,0), allowing us to visualize the spatial extent of the system consistently in both angular and comoving coordinates. Positions are shown in right ascension–declination (bottom/left axes) and in the comoving transversal plane (top/right axes). The central halo is marked with a cyan “$\times$”. Within the assembly history, we identify SFGs, LIRGs, ULIRGs, and HyLIRGs, marked respectively by light-blue, yellow, red, and green circles. Percentages in each panel report the fraction of galaxies in that snapshot belonging to each subset, computed relative to the total number of galaxies participating at that stage (star-forming + quiescent). Each panel also reports the mean redshift $\overline{z}$ and $\Delta z \equiv z_{\max}-z_{\min}$.}
  \label{fig:rich_cluster_evol}
\end{figure*}

Figure~\ref{fig:rich_cluster_evol} shows the assembly history of the richest cluster identified in our lightcone at $z=0.576$, according to the definition adopted in this work for a rich cluster (i.e., a bound system containing more than 50 member halos) displayed at several evolutionary stages. Although the system already satisfies our cluster definition at this redshift, its assembly is still ongoing, with some member halos remaining outside the virial radius of the main halo. The assembly history shown here includes the complete population of galaxies participating in this process, both quiescent and star-forming. Additionally, we highlight the luminous infrared galaxy population throughout the evolution of the system. Each snapshot presents the projected distribution of halos in right ascension–declination and in the equivalent comoving transversal plane.

At $z = 5.6$, the system appears as a compact proto-cluster with 116 members, mostly low-mass halos ($10^{11}$–$10^{12}$ M$_{\sun}$). The galaxy population is dominated by quiescent systems ($\sim68$ per cent), with star-forming galaxies making up the remaining $\sim32$ per cent, including a fraction of LIRGs ($\sim22$ per cent) and ULIRGs ($\sim9$ per cent). By $z = 3.1$, the structure has grown substantially in membership, reaching a total of 289 halos. The mass of the central halo increases from $1.9\times10^{12}$ $\mathrm{M}_{\sun}$ to $3.6\times10^{13}$ M$_{\sun}$. The infrared-bright population also becomes more prominent: $\sim23$ per cent LIRGs, $\sim14$ per cent ULIRGs, and the first HyLIRGs ($\sim1$ per cent) appear, reflecting an enhanced phase of star formation. At $z \sim 1.9$, the system reaches the maximum number of progenitor galaxies in its merger tree, with a total of 312 members, displaying a massive and extended structure and a peak in star-forming activity. The main halo has a mass of $2.7\times10^{14}$. By $z =1.0$, the system is more compact, with nearly half of the galaxies already quenched, although a residual population of infrared-bright galaxies persists. At this epoch the main halo reach a mass of $8.6\times10^{14}$. Finally, at $z=0.576$, the system consists of 122 cluster members, comprising the main halo and 121 associated subhalos. Each halo hosts a galaxy with a stellar mass of $M_\star \geq 10^{9.5}\,\mathrm{M}_{\sun}$, corresponding to the stellar mass resolution limit in our lightcone. The main halo attains its maximum virial mass of $2.0\times10^{15}\,\mathrm{M}_{\sun}$ at this redshift. At the same epoch, the merger tree identifies 71 additional progenitor halos that are not yet cluster members but will eventually assemble into the final cluster at $z=0$, resulting in a total of 193 progenitor halos at $z=0.576$. Although the system continues to evolve towards lower redshifts, its descendants fall outside the volume sampled by our lightcone. Therefore, its evolution cannot be followed to $z=0$ within the present dataset. At this epoch, the cluster enters a phase with little or no star formation activity, the cluster contains only one ULIRG galaxy in a total of 193 galaxies (0.5 per cent) involved in this stage of the assembly history.

\begin{figure*}
  \centering
  \includegraphics[width=\textwidth]{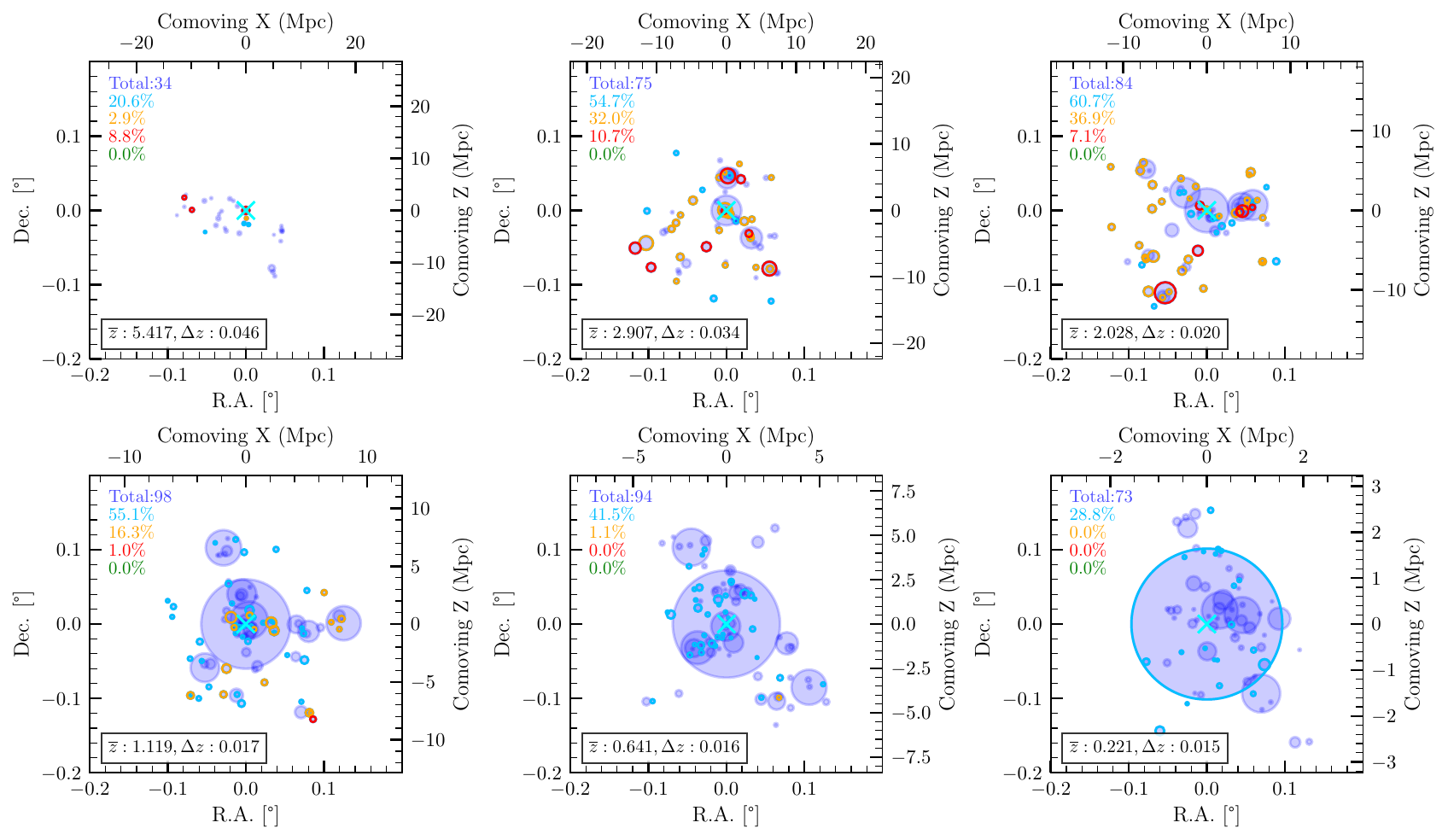}
  \caption{Same as Figure \ref{fig:rich_cluster_evol}, but for a more typical cluster containing 56 members and a main halo mass of $6.9\times10^{14}\,{\rm M}_{\sun}$ at $z=0.22$. In addition to its 56 members, 17 halos linked through the merger tree are identified as progenitors of the final cluster at $z=0$, giving a total of 73 progenitor halos associated with the system at this $z=0.22$.}
  \label{fig:typical_cluster_evol}
\end{figure*}

Figure \ref{fig:typical_cluster_evol} shows another example of assembly history, but for a typical rich cluster. Containing 56 members and a central halo with a total mass of $6.9\times10^{14}\;{\rm M}_{\sun}$ at $z=0.22$. In this case, a lower level of star formation activity is observed throughout all evolutionary stages. At early times ($z \sim 5.4$), the system is dominated by quiescent galaxies (80 per cent), with only a few LIRGs (2.9 per cent) and ULIRGs (8.8 per cent). The peak of star formation activity occurs between $z \sim 3$ and $z \sim 2$. By $z=1.1$, the star formation phase has mostly ended, with only one ULIRG remaining in the outskirts of the system (16 per cent). At $z \sim 0.6$, 40 per cent of the galaxies are still forming stars, but none exhibit high star formation rates ($>100\; \rm M_{\sun}$ yr$^{-1}$). Finally, at $z \sim 0.2$ the majority of galaxies are quenched (72 per cent), while, interestingly, the central galaxy is classified as star-forming.

\subsection{Contribution of proto-clusters to the SFRD}

We investigate the contribution of proto-clusters in our simulation to the cosmic $\mathrm{SFRD}$. In our analysis, we identify both poor and rich clusters and impose a halo mass threshold of $M_{200} > 10^{14}\;{\rm M}_{\sun}$, where $M_{200}$ is the mass enclosed within $R_{200}$, inside which the mean overdensity is 200 times the critical density. We follow the backward evolution of these systems across cosmic time to estimate their contribution to the total $\mathrm{SFRD}$.

\begin{figure}
  \centering
  \begin{subfigure}[b]{1\columnwidth}
    \includegraphics[width=\columnwidth]{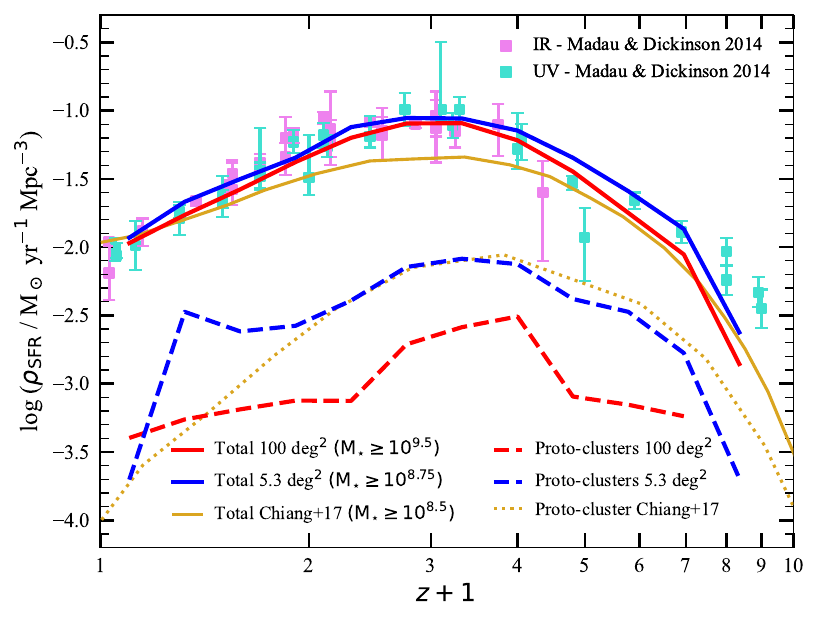}
  \end{subfigure}
  
  \vspace{0.em} 

    \begin{subfigure}[b]{1\columnwidth}
    \includegraphics[width=\columnwidth]{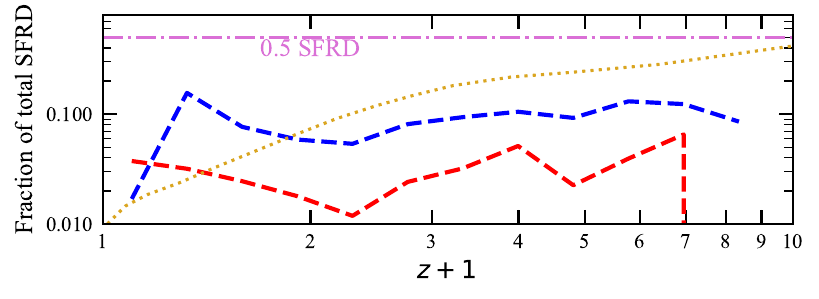}
  \end{subfigure}
  \caption{\textbf{Top panel:} Total cosmic $\mathrm{SFRD}$ predicted by our 100 square degrees mock lightcone (solid red line) and by the 5.3 square degrees catalogue of \citetalias{nava-moreno+24} (solid blue line), compared with the results of \citet{chiang+2017} (solid golden line). The corresponding contributions from proto-clusters with $M_{200c}>10^{14}\,{\rm M}_{\sun}$ are shown with dashed-dotted and dotted lines for each dataset. Observational constraints from \citet{madau_dickinson2014} (UV and IR) are also shown.  \textbf{Bottom panel:} Fraction of the total $\mathrm{SFRD}$ contributed by proto-clusters in each catalogue, with lines following their corresponding datasets. The magenta horizontal line marks the 50 per cent level for reference.}
  \label{fig:sfrd_contribution}
\end{figure}

Given that our mock catalogue is limited to a stellar mass of $M_{\star}=10^{9.5}\,{\rm M}_{\sun}$, we also assess how the stellar mass resolution affects the inferred contribution of proto-clusters to the cosmic $\mathrm{SFRD}$. To quantify this effect, we apply the same cluster identification methodology to the higher resolution mock catalogue GARDENS-Deep of \citetalias{nava-moreno+24}, which reaches a stellar mass resolution of $M_{\star}=10^{8.75}\,{\rm M}_{\sun}$. Figure~\ref{fig:sfrd_contribution} shows the total $\mathrm{SFRD}$ and the contribution from proto-clusters for both mock catalogues. The higher resolution catalogue systematically yields larger $\mathrm{SFRD}$ values, indicating that a fraction of the star formation in low mass galaxies is missed in the 100~square degree lightcone. This trend is reflected in the proto-cluster contribution as well.

We compare our results with those of \citet{chiang+2017}, who select $M_{200} > 10^{14}\;{\rm M}_{\sun}$ galaxy clusters at $z=0$ from the Millennium $N$–body simulation ($L = 500\,h^{-1}\,\mathrm{Mpc}$) with $M_{200} > 10^{14}\,{\rm M}{_{\sun}}$ and then trace their merger trees back in time to study their proto–cluster progenitors. Their catalogue reaches a stellar mass resolution of $M_{\star}=10^{8.5}\,{\rm M}_{\sun}$. For comparison, Figure~\ref{fig:sfrd_contribution} also includes the total $\mathrm{SFRD}$ and proto-cluster contribution reported by \citet{chiang+2017}.

When focusing on the proto–cluster contribution to the $\mathrm{SFRD}$, we find relative contributions of $6.5$ per cent in this work and $\sim12$ per cent in \citetalias{nava-moreno+24} at $z>5$. In comparison, \citet{chiang+2017} report significantly higher values, reaching $\sim30$ per cent at similar redshifts, although their total $\mathrm{SFRD}$ is systematically lower than that measured in our mock catalogue and in \citetalias{nava-moreno+24}. On average, the total $\mathrm{SFRD}$ in \citet{chiang+2017} is lower by $\sim0.20$ dex, corresponding to a factor of $\sim0.64$ compared to \citetalias{nava-moreno+24}. Taking this into account reduces their reported contribution from $\sim30$ per cent to $\sim20$ per cent, partially reducing the difference with our results. However, differences in stellar mass resolution may also play a role, as shown by our results, since they affect the number of low–mass galaxies in proto–cluster environments and therefore their contribution to the $\mathrm{SFRD}$. At lower redshifts ($z<1$), our predictions should be interpreted with caution owing to the limited statistics at the narrow end of our lightcone.

\section{Luminous Infrared Galaxy Populations in Cluster Environments} \label{sec:cluster_analysis}

In this section, we characterize the population of bright infrared galaxies across different environments (clusters and groups) and track how the fraction of infrared bright galaxies residing in these structures evolves during their assembly. We also analyze how the projected and three-dimensional physical sizes of clusters evolve across cosmic time. Finally, we examine the detectability of galaxy pairs at the submillimetre wavelengths, and quantify the fraction of systems that can be identified as pairs based on their individual submillimetre fluxes.

\subsection{Characterizing the Infrared Component of Galaxy Clusters}

\begin{figure*}
  \centering
  \begin{subfigure}[b]{0.32\textwidth}
    \includegraphics[width=\linewidth]{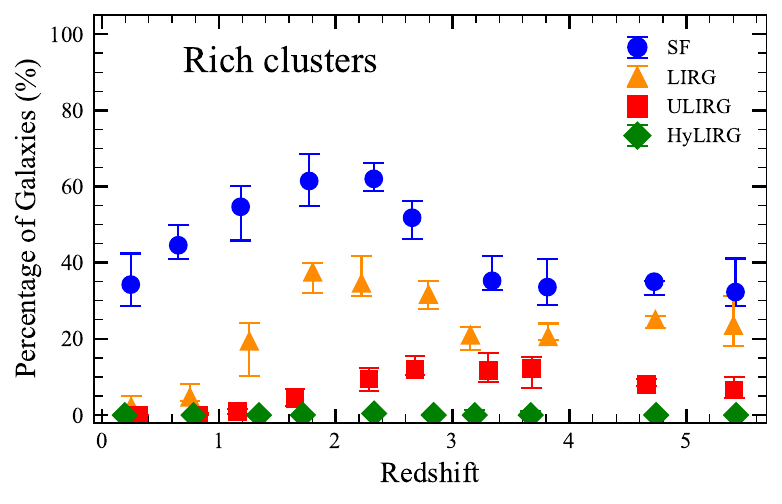}
  \end{subfigure}
  \hfill
  \begin{subfigure}[b]{0.32\textwidth}
    \includegraphics[width=\linewidth]{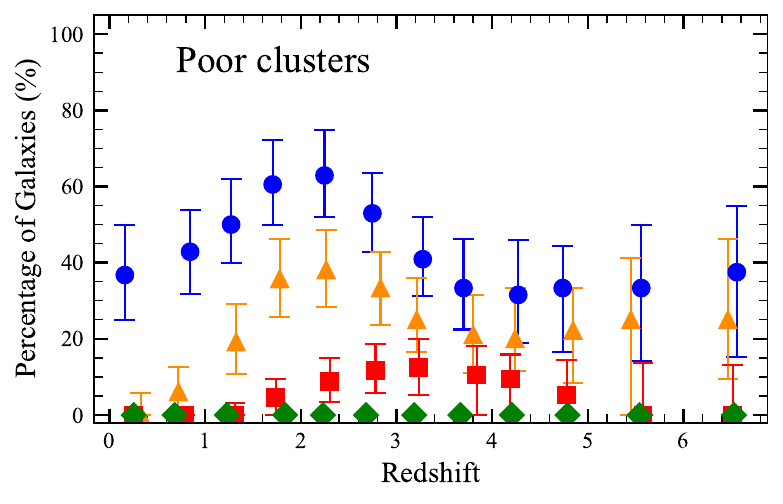}
  \end{subfigure}
  \hfill
  \begin{subfigure}[b]{0.32\textwidth}
    \includegraphics[width=\linewidth]{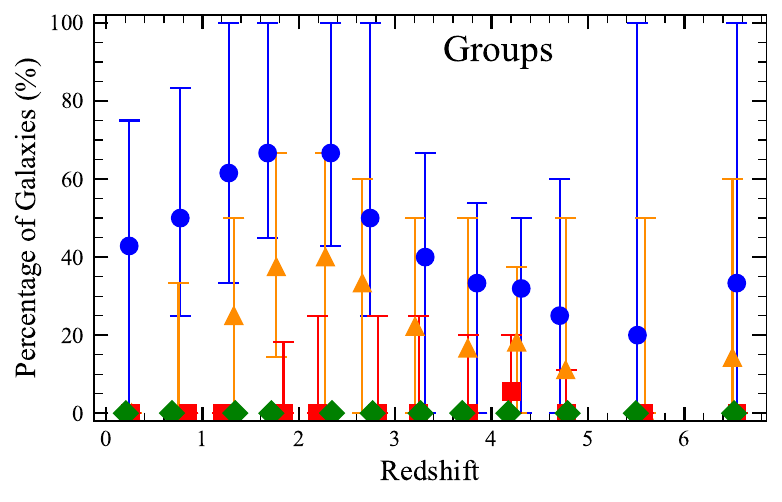}
  \end{subfigure}
  \caption{Median percentage contribution of star-forming galaxies (blue) and their infrared-bright subpopulations, including LIRGs (yellow), ULIRGs (red), and HyLIRGs (green), to the assembly histories of groups, poor clusters, and rich clusters shown in the three panels. The contributions are computed relative to the total number of galaxies participating in each category (including both star-forming and quiescent galaxies) within each redshift bin. The bars represent the 16th–84th percentile range, illustrating the scatter in the distribution.}
  \label{fig:porcentaje_poblaciones}
\end{figure*}

To investigate how different star-forming and IR galaxy populations contribute across the assembly history of each category (clusters, poor clusters, and groups), we identify galaxies within each history as star-forming galaxies, and among them LIRGs, ULIRGs, and HyLIRGs, and further bin them in redshift. Figure \ref{fig:porcentaje_poblaciones} shows the median fraction (in per cent) of each population within each category as a function of redshift, with the error bars representing the 16th–84th percentile range of the scatter in the distribution. The percentages are computed relative to the total number of galaxies (including both star-forming and quiescent galaxies) participating in the assembly history of each redshift bin. We find that star-forming galaxies contributing to the assembly history of rich and poor cluster category contribute about 35 per cent at $z<0.5$, with the fraction steadily increasing with redshift and reaching a peak of $\sim60$ per cent at $z\sim2$. Beyond this point, the contribution gradually declines to $\sim30$ per cent, forming a tail that extends to $z\sim5.5$, beyond which no further information is available due to the discretization of redshifts in the lightcone replication. For the assembly of groups, the trend is similar to that of clusters: the contribution rises with redshift and peaks at $\sim65$ per cent around $z\sim2$, where it starts to decline until $z\sim6$, where, the percentage increases from 20 per cent to 33 per cent but with a considerably large scatter.

LIRGs contribute between 20–40 per cent across all categories, with a peak around $z \sim 2$. Their contribution is relatively stable in the assembly histories of systems that evolve into clusters, while in those evolving into groups it shows larger scatter. ULIRGs remain subdominant, contributing about 5–12 per cent at $z>1.5$ in cluster assembly histories, but their fraction drops to nearly zero in group progenitors except for a brief increase around $z\sim4$. 

Across the full lightcone, we identify a total of 5,724 HyLIRGs, of which 25 per cent are hosted by isolated galaxies, 43 per cent by pairs, 32 per cent by groups, 0.2 per cent by poor clusters, and none by rich clusters. Therefore, the HyLIRG population is dominated by pair and group environments. When normalized by the total galaxy population in each environment, however, HyLIRGs remain extremely rare, representing only 0.02 per cent, 0.4 per cent, 0.8 per cent, and 0.23 per cent of the isolated galaxy, pair, group and poor cluster, respectively. When tracing the assembly histories of cluster systems, we find that 20 HyLIRGs are associated with protoclusters of future rich clusters and 490 with protoclusters of future poor clusters, corresponding to 9 per cent of the total HyLIRG population in the lightcone. Only 11 per cent of the protoclusters of future rich clusters and 3 per cent of the protoclusters of future poor clusters host between one and three HyLIRGs, indicating that HyLIRGs are present in only a minority of protocluster systems.

\subsection{Radial Distribution during Cluster Assembly}

Once we separate the LIRG and ULIRG populations in the assembly histories of poor and rich clusters, we study their spatial distribution within the forming structures, whether these luminous galaxies are located in the inner regions, at the outskirts, or spread throughout the proto-cluster volume. To quantify this, we define the size of each system in terms of its maximum 3D comoving distance, measured as the maximum distance between the central halo (selected as the most massive halo in each snapshot) and its most distant satellite. We also compute the maximum projected angular radius on the sky, which provides an observational proxy for the apparent size of the forming structures proto-clusters.

\begin{figure*}
    \centering
    \includegraphics[width=0.9\textwidth]{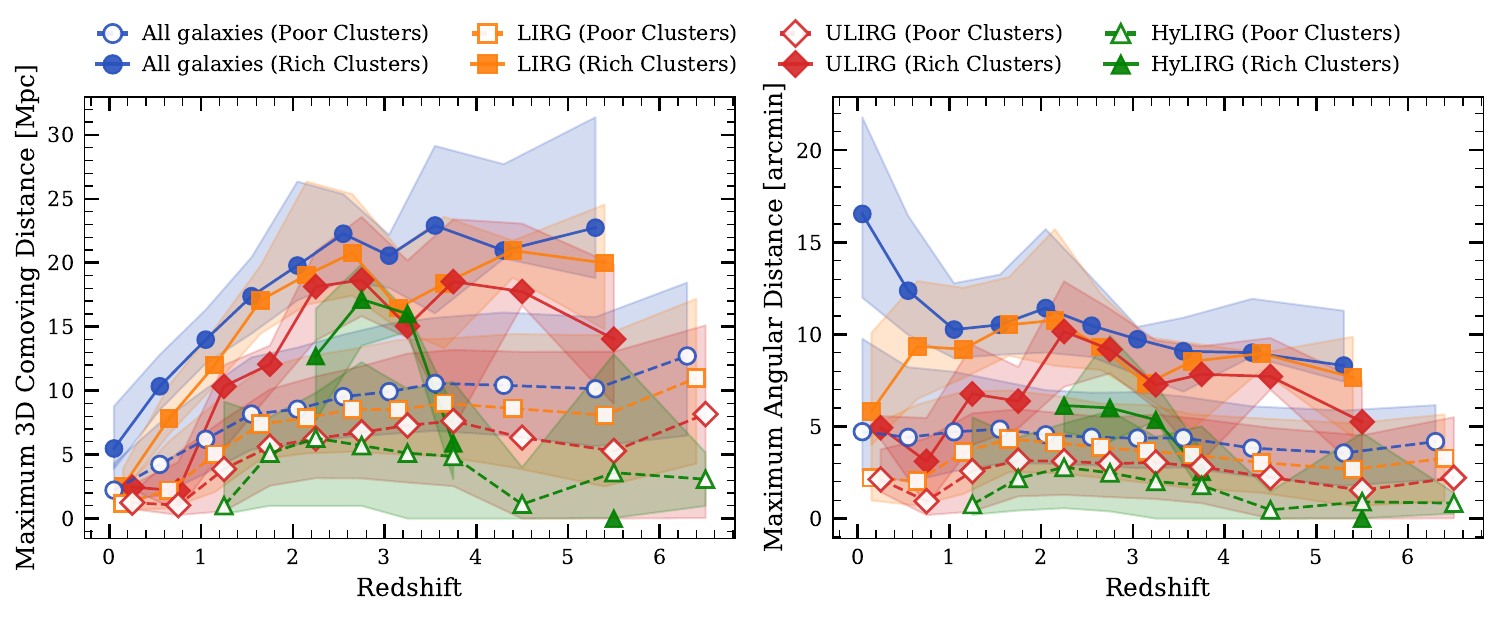}
    \caption{Median of the maximum distance between the central halo of each proto-cluster and its most distant satellite galaxy at each stage of its evolution, shown as a function of redshift. This quantity traces the spatial extent of the assembling system for rich clusters ($>$50 members, solid symbols) and poor clusters (10–50 members, open symbols). \textbf{Left panel:} Results for the three-dimensional comoving distance. \textbf{Right panel:} Results for the projected angular distance. Colors indicate different galaxy populations: the full proto-cluster population (star-forming and quiescent galaxies; blue) and infrared-bright subsamples of the star-forming population, namely LIRGs, ULIRGs, and HyLIRGs (orange, red, and green, respectively). Shaded bands represent the 16th--84th percentile range of the distributions.}
    \label{fig:max_dist}
\end{figure*}

Figure~\ref{fig:max_dist} presents the median 3D comoving size, as a function of redshift for rich and poor clusters. For this analysis, we consider four distinct cases: (i) the total system, including both star-forming and quiescent galaxies; (ii) the system traced by LIRGs; (iii) the system traced by ULIRGs and (iv) the system traced by HyLIRGs. This allows us to simultaneously characterize the overall growth of proto-cluster size and to investigate the spatial distribution of the most infrared-luminous galaxy populations. We also examine whether subdividing the sample by halo mass affects the results; however, no systematic differences were found. We therefore present the results without applying any additional mass selection.

The results show that the progenitors of rich clusters exhibit a median maximum comoving size of $\sim22$ Mpc at $z \sim 5.5$, decreasing mildly to $\sim20$ Mpc by $z \sim 2$, and more rapidly to $\sim5.3$ Mpc at $z \sim 0$. This evolution indicates that proto-clusters initially span large regions that progressively contract in comoving coordinates, reflecting their gradual decoupling from the expansion of the Universe as gravitational collapse becomes dominant.

In physical units, the trend is more intuitive. The median maximum size of rich clusters progenitors increases from $\sim3.4$ Mpc at $z \sim 5.5$ to $\sim6.6$ Mpc at $z \sim 2$, and then decreases to $\sim5.3$ Mpc at $z \sim 0$. This behaviour shows that proto-clusters grow in size due to ongoing accretion, reach a maximum extent around $z \sim 2$, and subsequently collapse into compact, gravitationally bound systems.

LIRGs are slightly more centrally concentrated, from 20 Mpc at $z \sim 5.5$ to 2.4 Mpc at $z < 0.2$, while ULIRGs are even more concentrated, ranging from 14 Mpc at $z \sim 5.5$ to $\sim2$ Mpc by $z < 1$, indicating that the most actively star-forming galaxies preferentially inhabit the densest regions. HyLIRGs appear to be even more centrally concentrated than ULIRGs, with their distances spanning from $\sim$6 Mpc at $z \sim 3.7$ (where ULIRGs extend to $\sim$18.5 Mpc) to about 17 Mpc at $z \sim 2.7$ (where ULIRGs reach $\sim$18.7 Mpc). However, these HyLIRG distances should be interpreted with caution, as the statistics in this population are very limited. At $z\sim5.5$, the HyLIRG bin in rich cluster progenitors correspond to a single object, resulting in no measurable dispersion.

In poor clusters, the median maximum comoving distance declines from $\sim13$ Mpc at $z \sim 6.3$ to $\sim2.2$ Mpc at $z \sim 0$, consistent with their smaller sizes and lower membership relative to rich systems. In physical units, this corresponds to $\sim1.8$ Mpc at high redshift, increasing to $\sim3$–4 Mpc at intermediate epochs, and then decreasing to $\sim2.2$ Mpc by $z \sim 0$. Despite following the same overall behaviour, poor cluster progenitors evolve more smoothly than rich ones, with a more gradual accretion phase and a less pronounced contraction, in line with their lower masses and shallower potential wells.

For HyLIRGs in poor clusters, the maximum comoving distances increase from $\lesssim$3.5 Mpc at $z\sim5.5$ to about 6 Mpc at $z\sim2.5$, and finally decrease to $\sim$0.9 Mpc by $z\sim1.2$. Over the same redshift range, ULIRGs extend out to $\sim$5.3 Mpc at $z\sim5.5$, $\sim$5 Mpc at $z\sim2.5$, and $\sim$4 Mpc at $z\sim1.2$, indicating that HyLIRGs are typically more centrally concentrated than ULIRGs.

To assess the robustness of our proto-cluster size definition, we also estimated the size using the mean distance from the central halo to the three and four most distant halos (or galaxies) at each snapshot, instead of the single most distant object. These alternative definitions systematically yield slightly smaller sizes than the maximum distance estimate. For the overall galaxy population, the difference is only 10–13 per cent. The differences become larger for the LIRG and ULIRG populations, reaching up to $\sim40$ per cent in the most extreme cases. This behaviour is likely due to the low abundance of these galaxies, making the distance estimates more sensitive to the spatial distribution of individual galaxies within the proto-cluster.  

In terms of projected angular size in the sky, our rich cluster progenitors appear larger at low redshift with median sizes of 16.5 arcmin at $z\sim0$ and gradually decrease decreasing to 8 arcmin while increasing redshift as redshift increases, consistent with the expected behaviour set by the angular diameter distance. Poor cluster progenitors show much smaller apparent sizes and weak evolution with redshift: its median angular size is 4.8 arcmin at $z\sim 0$ and 4.2 at $z\sim6.3$. This reflects their intrinsically smaller comoving sizes, which project to nearly constant angular scales across cosmic time. For rich and poor clusters at $z\sim0$, ULIRGs remain more centrally concentrated than the total galaxy population, with median distances of 5 and 2.2 arcmin, respectively, and similar values of 5.2 and 2.3 arcmin at $z\sim5$–6. However, at $z\sim2.2$ the median distance in the rich cluster progenitors increases to $\sim$10 arcmin, while in the poor cluster progenitors it remains nearly constant. HyLIRGs exhibit a more compact distribution: at $z\sim2.2$ their median angular distances are $\sim$6 arcmin in rich clusters and $\sim$2.7 arcmin in poor clusters, indicating that they preferentially trace the innermost regions of their host systems.

We also show the 16th to 84th percentile range of the distributions, providing a measure of the scatter between systems at fixed redshift. This information complements the median trends by showing not only the typical growth of clusters but also the degree of variation in their assembly histories, which tends to be larger at higher redshifts when structures are still forming.

Both the comoving and the angular distances provide complementary views of cluster sizes, which are highly valuable for the design of future surveys, helping to optimize angular resolution, field coverage, and the identification of member galaxies across redshift.

\begin{figure*}
  \centering
  \begin{subfigure}[b]{.7\textwidth}
    \includegraphics[width=\linewidth]{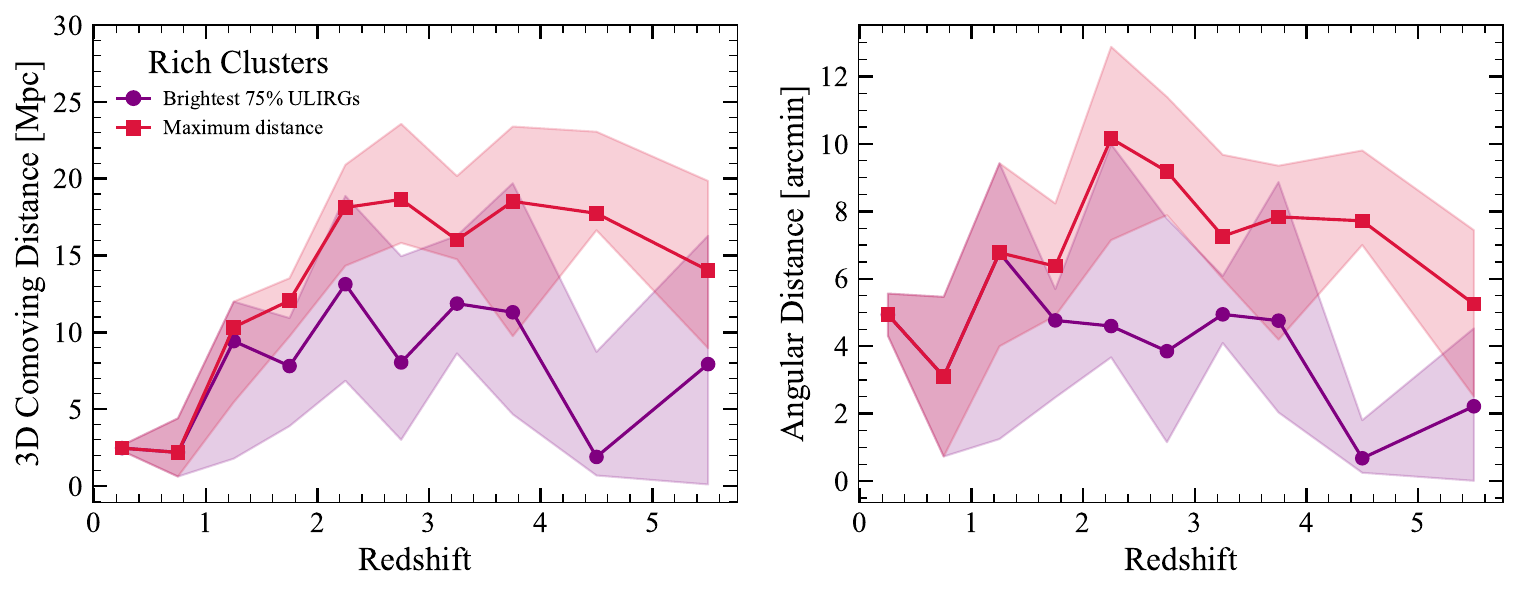}
  \end{subfigure}
  \vspace{0.0em} 
  \begin{subfigure}[b]{0.7\textwidth}
    \includegraphics[width=\linewidth]{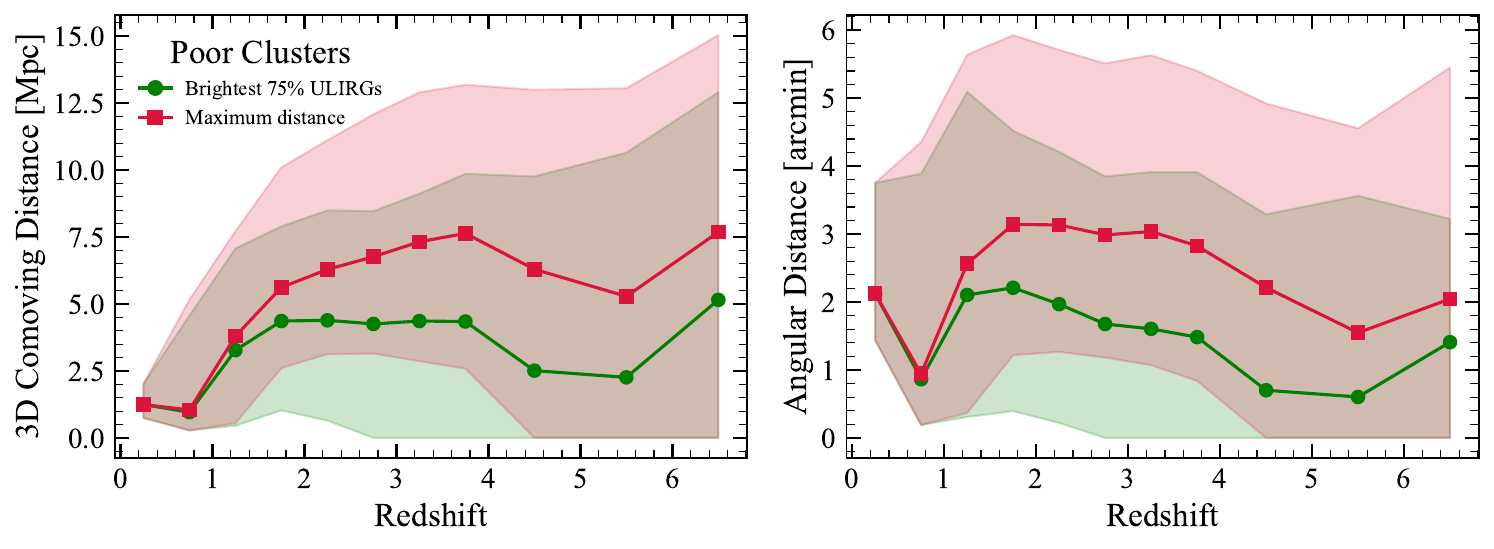}
  \end{subfigure}
    \caption{Median 3D distance between the proto-cluster centre and the 75th-percentile ULIRG (purple and green lines), compared with the distance to the most distant ULIRG (red lines), for proto-clusters whose descendants are rich (top panel) and poor (bottom panel) clusters. Shaded regions indicate the 16th–84th percentile range. The figure also shows, on the right panels, the equivalent angular distance on the sky at each redshift.}
    \label{fig:dist_75_ulirg}
\end{figure*}

Having characterized the overall spatial extent of the cluster galaxy population, we now focus on the distribution of the more luminous ULIRGs within these systems. We also measure the 3D comoving distance between the cluster center and the ULIRGs at the 75th percentile of the infrared-luminosity distribution at each evolutionary stage (after ranking ULIRGs from faintest to brightest) to assess how these galaxies are spatially distributed. Figure \ref{fig:dist_75_ulirg} compares this distance with that of the most distant ULIRG galaxy, as shown in Figure \ref{fig:max_dist} for proto-clusters that evolve in rich and poor clusters. At $z > 1.5$, the 75th percentile ULIRGs tends to be more concentrated towards the cluster center than towards the outskirts when compared with the overall ULIRG population. The median maximum comoving distance reached by the 75th percentile ULIRG is 13 and 7.5 Mpc for rich and poor clusters, respectively. At $z < 1.5$, the convergence between the distance to the 75th percentile ULIRG and the total ULIRG population can be explained by two complementary effects. First, the number of ULIRGs drops sharply at low redshift, so the percentile estimate becomes poorly constrained: with only a few galaxies, the 75th percentile ULIRG value often coincides either with the central or the most distant ULIRG. Second, clusters at these epochs are already more dynamically relaxed, and the few remaining ULIRGs do not occupy a preferential location within the system. As a result, their spatial distribution is indistinguishable from that of the overall galaxy population, leading to the observed similarity between both distance measures. Figure \ref{fig:dist_75_ulirg} also shows the equivalent to the angular distance.

\subsection{Galaxy pairs}

In terms of abundance, galaxy pairs are the second most common population in our mock catalogue, after isolated galaxies. However, unlike clusters or groups, pairs do not represent an environment with a well-defined evolutionary history, since they only consist of two galaxies. For this reason, we do not attempt to study their evolution, and instead focus on their statistical properties and on how many of them can be detected in the submillimetre bands.

In our 100 square degrees mock survey we identify 585,334 galaxy pairs. Using our star-forming selection (Subsection~\ref{subsec:2.2}), we classify each pair according to whether the central and the satellite galaxies are star-forming or not. Here, central and satellite galaxies are defined according to the halo–subhalo classification obtained in the simulation (i.e. based on the halo/subhalo identifiers). Figure \ref{fig:percent_classification} shows the evolution of the pair population as a function of redshift, presenting the fraction of pairs relative to the total number of galaxies in each redshift bin in our 100 square degrees mock redshift survey. 

Overall, 42.5 per cent of all pairs consist of two star-forming galaxies, making this the most common configuration. Among the remaining systems, 24.7 per cent correspond to pairs in which only the central galaxy is star-forming, 18.2 per cent to pairs in which only the satellite is star-forming, and 14.6 per cent to pairs in which neither galaxy is star-forming. Therefore, nearly half of the pair population consists of systems in which both galaxies are star-forming and are potentially observable as galaxy pairs at submillimetre wavelengths. The actual fraction of observable pairs will depend on the sensitivity and angular resolution of the instrument. In Section~\ref{sec:detectable_pairs_toltec}, we assess this using the capabilities of TolTEC.

\section{Predictions for the LSS \TolTEC\ survey}\label{sec:toltec_predictions}

In this section, we present predictions for the source counts, redshift distribution, and galaxy pairs detectable in the LSS TolTEC survey, as well as the expected proto-cluster population. The survey will cover 60 square degrees and is designed to reach the ULIRG regime ($L_{\rm IR} \geq 10^{12}\,\rm L_{\sun}$) at submillimetre wavelengths. These predictions serve as a benchmark for interpreting forthcoming TolTEC observations.

\subsection{Predicted source counts}

Based on the nominal survey depths of the 60\,deg$^2$ LSS TolTEC survey, which reach $1\sigma$ sensitivities of 0.25, 0.18, and 0.12\,mJy\,beam$^{-1}$ at 1.1, 1.4, and 2.0\,mm respectively, we predict that 104,000 (1.1~mm), 50,000 (1.4~mm), and 11,000 (2.0~mm) sources lie above the corresponding 4$\sigma$ detection limits. In terms of infrared luminosity, we estimate that the flux limited samples contain about 90,000 ULIRGs and 3,000 HyLIRGs at 1.1~mm; 44,000 ULIRGs and 3,300 HyLIRGs at 1.4~mm; and 8,900 ULIRGs and 1,500 HyLIRGs at 2.0~mm. Table~\ref{tab:toltec_amplified} lists the number of galaxies expected at each TolTEC wavelength and depth, including the counts of lensed galaxies above different amplification limits over the 60 square degree survey area.

\begin{table}
\centering
\caption{Predicted source counts in the 60\,deg$^2$ LSS TolTEC survey at 1.1, 1.4, and 2.0\,mm. The total is rounded to the nearest thousand and includes both unlensed ($\mu = 1$) and gravitationally lensed populations, with the latter further divided by magnification thresholds, where the minimum magnification considered is $\mu > 1.2$.}
\begin{tabular}{lccc}
\hline
\multicolumn{4}{c}{60\,deg$^2$ LSS TolTEC survey} \\[3pt]
\hline
Observed wavelength (mm) & 1.1 & 1.4 & 2.0 \\
\hline
Total detected sources & 104,000 & 50,000 & 11,000 \\
Unlensed sources ($\mu = 1$) & 95,763 & 44,791 & 9,171 \\
Lensed sources ($\mu > 1.2$) & 9,047 & 5,254 & 1,640 \\
Lensed sources ($\mu > 10$) & 398 & 306 & 176 \\
Lensed sources ($\mu > 20$) & 136 & 119 & 80 \\
Lensed sources ($\mu > 50$) & 31 & 30 & 24 \\
\hline
\end{tabular}
\label{tab:toltec_amplified}
\end{table}

\subsection{Redshift distribution}

\begin{figure}
    \centering
    \includegraphics[width=1\columnwidth]{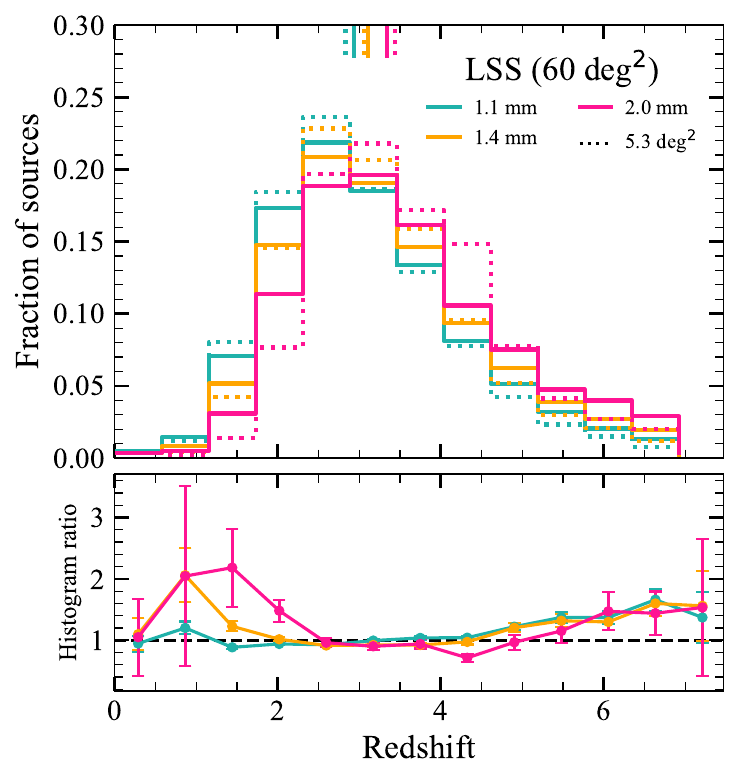}
    \caption{Redshift distributions for the TolTEC LSS survey (60 square degrees) at 1.1, 1.4, and 2.0 mm, assuming 4$\sigma$ limits of 1.0, 0.72, and 0.48 mJy\,beam$^{-1}$. Dotted histograms show results from \citetalias{nava-moreno+24} (5.3 square degrees) with the same flux selection. Vertical lines indicate the median redshift in each band. The lower panel shows the ratio of the normalized histograms, with propagated Poisson uncertainties derived from the source counts in each redshift bin.}
    \label{fig:zdist_toltec}
\end{figure}

Figure~\ref{fig:zdist_toltec} presents the predicted redshift distribution for the TolTEC LSS survey over 60 square degrees at 1.1, 1.4, and 2.0 mm considering galaxies above the corresponding 4$\sigma$ detection thresholds (1.0, 0.72, and 0.48 mJy\,beam$^{-1}$, respectively). The median redshifts are 2.9, 3.1 and 3.3 for 1.1, 1.4, and 2.0 mm, showing a systematic shift towards higher redshift at longer wavelengths. We compare these values with the redshift distributions obtained over the 5.3 square degrees area presented in \citetalias{nava-moreno+24}, where median redshifts of 2.8, 3.1, and 3.4 are derived using the same flux cuts at each wavelength. The median redshifts and the overall shapes of the distributions are consistent despite differences in survey area, stellar mass resolution, and the modelling of the infrared properties. The lower panel shows the ratio of the normalized redshift distributions (60 square degrees relative to that in the 5.3 square degrees), with propagated Poisson uncertainties derived from the source counts in each redshift bin. The ratio remains close to unity over the redshift range containing the majority of the sources, while the larger differences at low and high redshift are associated with a lower number of galaxies in those bins. The observed shift of the median redshift towards higher values at longer wavelengths is consistent with the findings of \citet{zavala+2014}, who showed that this behaviour is primarily driven by selection effects related to the observing wavelength and survey depth, rather than intrinsic differences in the galaxy population.

\subsection{Detectability in galaxy pairs} \label{sec:detectable_pairs_toltec}

To estimate how many galaxy pairs can be observed in the LSS TolTEC survey, we consider the 4$\sigma$ detection limits and angular resolution at each observing band (1.1, 1.4, and 2.0 mm), using their angular separations to determine whether the pair members can be detected as distinct sources or would appear blended.

We focus on the 248,766 galaxy pairs in which both galaxies are star-forming and identify those in which both members are detectable at each band. We find that 0.42 per cent, 0.11 per cent, and 0.01 per cent of the pairs are detectable at 1.1, 1.4, and 2.0 mm, respectively. Interestingly, these systems are found at relatively higher redshifts than the general population probed by the TolTEC LSS survey (Figure \ref{fig:zdist_toltec}), with median values of 3.2, 3.5, and 3.9, respectively.

However, a significant fraction of these systems are not resolved by the TolTEC beam. At 1.1 mm, where the beam size is 5 arcsec, only 36 per cent of the detectable pairs are resolved, while the remaining 64 per cent would appear blended. This effect becomes more pronounced at longer wavelengths: at 1.4 mm (FWHM = 6.3 arcsec), only 16 per cent of the pairs are resolved, and at 2.0 mm (FWHM = 9.5 arcsec), this fraction drops to $\sim$10 per cent, indicating that most detectable systems would be observed as single blended sources, particularly at longer wavelengths.

\begin{figure}
  \centering
    \subfloat{\includegraphics[width=\columnwidth]{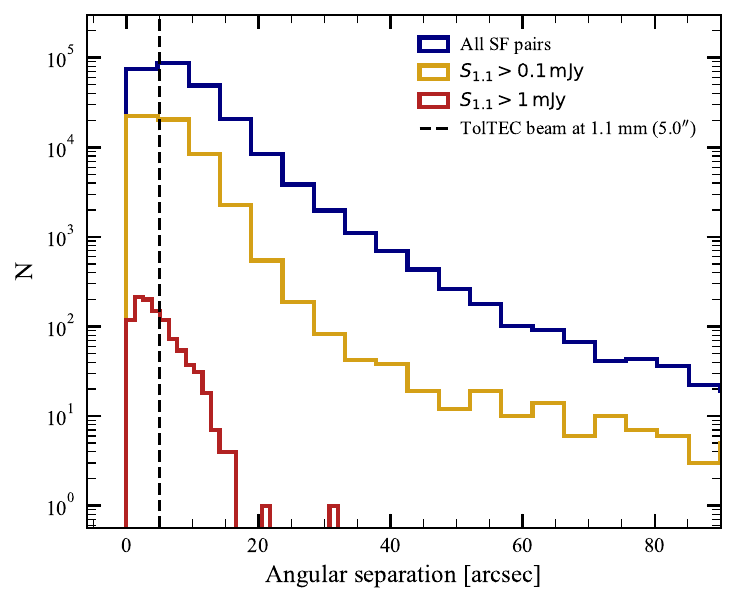}}    
  \caption{Distribution of angular distances for galaxy pairs in which both members are star-forming. N denotes the number of pairs. The blue histogram shows all pairs (without any flux limit), while the yellow and red histograms correspond to systems where both galaxies satisfy $S_{1.1} > 0.1$ mJy and $S_{1.1} > 1$ mJy, respectively. The vertical dashed line indicates the TolTEC beam at 1.1 mm (5 arcsec).}
  \label{fig:distance_pairs}
\end{figure}

Figure~\ref{fig:distance_pairs} shows the distribution of angular distances for galaxy pairs in which both members are star-forming, and subsamples where both galaxies satisfy $S_{1.1} > 0.1$ mJy and $S_{1.1} > 1$ mJy. We find that while the full population spans a broad range of separations, the detectable pairs are strongly biased towards small angular distances. In particular, a large fraction of the brightest systems lie at separations comparable to or below the TolTEC beam size at 1.1 mm, indicating that many of them would be observed as blended sources. This shows that blending is a dominant effect in the context of galaxy pairs and contributes to the overestimation of flux densities in observed systems.

\subsection{Clusters and Proto-clusters}

Over a sky area of 60 square degrees, equivalent to the expected area of the TolTEC LSS survey, we identify 19 rich clusters and 2,353 poor clusters. Due to the lightcone construction, different evolutionary stages of a given cluster traced backwards may appear in different box replications as a result of the snapshot lightcone intersection. We therefore count all proto-cluster occurrences within the lightcone. Using this approach, we find a total of 72 proto-clusters associated with rich clusters, of which 48 are located at redshift $z \geq 2$ and 19 at $z \geq 4$. For poor clusters, we identify 7,220 proto-clusters, with 5,313 at $z \geq 2$ and 2,254 at $z \geq 4$.

\begin{figure}
    \centering
    \includegraphics[width=1\columnwidth]{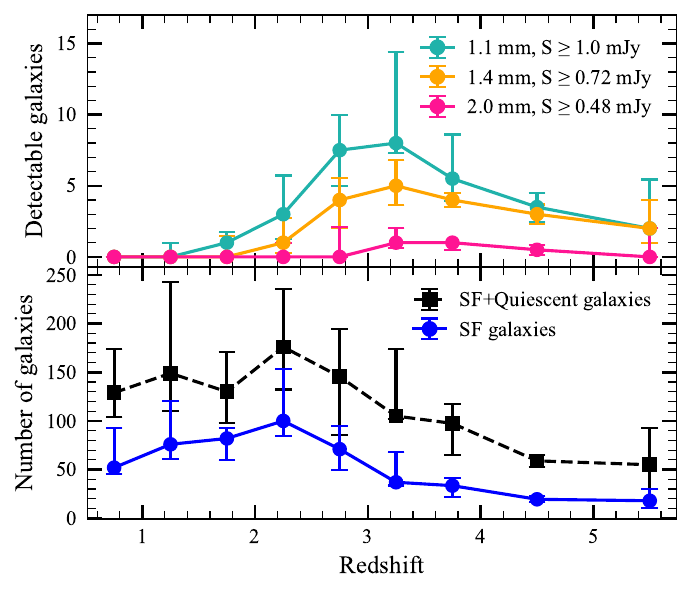}
    \caption{Median number of galaxies detectable in proto-cluster as a function of redshift at the three TolTEC bands (1.1, 1.4, and 2.0 mm in green, yellow and pink symbols), assuming the LSS 4$\sigma$ detection limits. Error bars indicate the 16th–84th percentile range. The number of detectable galaxies increases with redshift, reaching a peak at $z \sim 3$–3.5, and declines towards higher redshift. The lower panel shows the median total number of galaxies associated with each proto-cluster (black squares), as well as the median number of star-forming galaxies (blue dots).}
    \label{fig:det_gal_toltec}
\end{figure}

Using the full assembly histories of rich clusters and applying the TolTEC LSS 4$\sigma$ detection limits, we estimate the median number of galaxies that TolTEC is expected to detect in each proto-cluster as a function of redshift. As shown in Figure~\ref{fig:det_gal_toltec}, the median number of detectable galaxies increases with redshift. At 1.1 mm, we find a median of 3 detectable galaxies (1 at 1.4 mm and 0 at 2.0 mm) at $z \sim 2.2$. This number rises to a peak of 8 galaxies (5 at 1.4 mm and 1 at 2.0 mm) at $z \sim 3.2$, and then declines to 2 galaxies (2 at 1.4 mm and 0 at 2.0 mm) by $z \sim 5.5$. Figure~\ref{fig:det_gal_toltec} also shows the median total number of galaxies associated with proto-clusters, including both star-forming and quiescent populations, as well as the median number of star-forming galaxies alone.

\begin{figure}
    \centering
    \includegraphics[width=0.9\columnwidth]{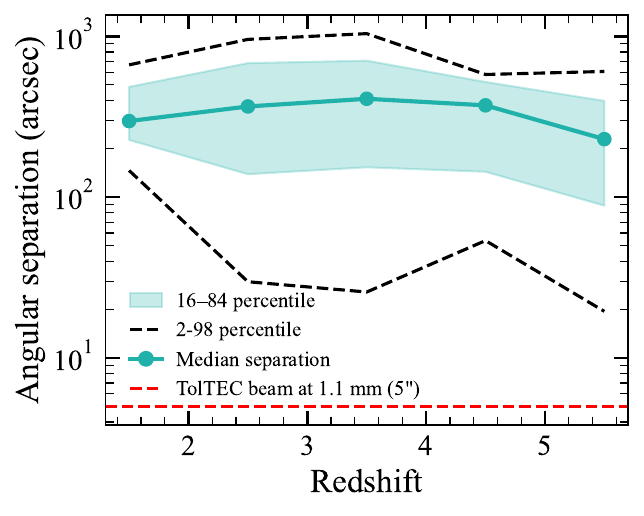}
    \caption{Median angular separation between detectable proto-cluster members as a function of redshift for systems that evolve into rich clusters. Only galaxies above the TolTEC LSS 4$\sigma$ detection limit at 1.1 mm are considered. The solid green line shows the median separation, while the shaded region indicates the 16–84th percentile range, and the dashed black lines correspond to the 2–98th percentiles. The horizontal dashed red line marks the TolTEC beam size at 1.1 mm (5 arcsec). Separations are computed between all unique pairs of detectable galaxies within each proto-cluster using their sky coordinates.}
    \label{fig:resolved_gal_toltec}
\end{figure}

However, the number of detectable galaxies alone does not determine whether these systems can be observationally identified as proto-clusters, as this also depends on the angular separation between their members relative to the instrumental resolution. To address this, we compute the angular separations between galaxies that are detectable according to the TolTEC LSS survey limits at 1.1\,mm, focusing on proto-clusters that evolve into rich clusters. We first select all galaxies across these systems whose fluxes exceed the detection threshold and then group them by their host proto-cluster at each redshift. Within each system, we calculate the separations between all unique pairs of detectable members using their sky coordinates (RA, Dec). This allows us to quantify how spatially separated these galaxies are and to assess whether they can be resolved given the TolTEC beam size at 1.1 mm (5 arcsec). Taking into account both the TolTEC detection limits and angular resolution at 1.1 mm, we find that detectable members are present in 40 out of the 48 proto-clusters identified in our 60 square degree mock survey. Figure~\ref{fig:resolved_gal_toltec} shows the median angular separation as a function of redshift, along with the 16–84th and 2–98th percentile ranges. We find that proto-cluster members become detectable only at redshifts $z \gtrsim 1.5$. At $z \sim 2$, the median separation between galaxies above the TolTEC detection limit is $\sim 300$ arcsec (5 arcmin). This value increases to nearly $\sim 400$ arcsec (6.6 arcmin) at $z\sim3.5$ and then decreases to $\sim 230$ arcsec (3.8 arcmin) at $z\sim3.5$. Overall, these separations are significantly larger than the TolTEC beam size.

To quantify the impact of blending in proto-clusters, we computed the fraction of detectable galaxy pairs with angular separations smaller than the TolTEC beam at 1.1 mm (5 arcsec). We find that only 5 out of 961 pairs ($\sim 0.5$ percent) fall below this threshold. This estimate considers only pairs within the same proto-cluster and does not include contamination from foreground or background galaxies. These results indicate that TolTEC will be able not only to detect but also to spatially resolve the vast majority of proto-cluster members at high redshift, with blending effects expected to be minimal under these conditions. 

Our results suggest that the main limitation of the TolTEC LSS survey for identifying proto-clusters is its sensitivity rather than its angular resolution. Although the majority of detectable galaxies are well resolved, only a small fraction of the proto-cluster galaxy population is expected to be detected, with the median number of detected members peaking at just 8 galaxies at 1.1~mm around $z\sim3.2$. Consequently, TolTEC will primarily probe the brightest dusty star-forming galaxy population within proto-clusters.

\section{Summary and conclusions} \label{sec:summary_conclusions}

We present GARDENS-Wide, a 100 square degree mock redshift survey of the DSFG population based on the MDPL2 dark matter cosmological simulation. This realization extends our previously published smaller-area catalogue GARDENS-Deep (\citetalias{nava-moreno+24}) to volumes large enough to contain the bright DSFGs population and sample rare environments such as massive clusters. The lightcone spans $0\le z\le7$ and reaches a stellar-mass resolution of $M_\star=10^{9.5}\,{\rm M_{\sun}}$, for more details see Table \ref{tab:comparing_simulations}. The catalogue is publicly available and forms the basis of the analysis presented here. 

We summarise our main results as follows:

\begin{enumerate}
 \item Analysis of the reconstructed assembly histories of clusters in our 100 square degree mock survey shows that proto-clusters contribute only a modest fraction (between 1–6 per cent) to the cosmic star-formation rate density, peaking at around 6 per cent at $z\simeq6$--7. In the higher stellar-mass resolution catalogue of \citetalias{nava-moreno+24}, this increases to approximately 12 per cent, indicating that the limited stellar-mass resolution underestimates the contribution from proto-clusters. Both are well below the $\sim30$ per cent reported by \citet{chiang+2017}; however, accounting for their lower total $\mathrm{SFRD}$ reduces this value to $\sim20$ per cent, reducing the discrepancy with our results. The remaining discrepancy may be related to differences in stellar–mass resolution.

\item Throughout the assembly of clusters and groups, star-forming galaxies dominate the build-up of structures, contributing about 35 per cent at low redshift and rising to 60--65 per cent at $z\sim2$. LIRGs represent a major component of the population, accounting for 20--40 per cent with a peak near $z\sim2$ in all environments. ULIRGs remain subdominant, reaching at most 5--12 per cent in clusters at $z>1.5$ and becoming negligible in groups, while HyLIRGs are extremely rare ($<0.5$ per cent) and do not play a significant role in the assembly of these systems.

\item In comoving units, Clusters experience strong contraction during their evolution. Rich cluster steadily decrease in size from $\sim$22 Mpc at $z \sim 5.5$ to $\sim$5.3 Mpc at $z \sim 0$, reflecting the increasing dominance of gravitational collapse over cosmic expansion. In physical units, the evolution is more intuitive: proto-clusters grow from $\sim$3.4 Mpc at $z \sim 5.5$ to a maximum of $\sim$6.6 Mpc at $z \sim 2$, driven by ongoing accretion, and then decrease to $\sim$5.3 Mpc at $z \sim 0$ as they become compact, virialized systems. In rich clusters, LIRGs are typically found at smaller comoving radii than the overall galaxy population, with radii decreasing from $\sim$20 Mpc at $z\sim5.5$ to 2.4 Mpc at $z<0.2$, while ULIRGs are even more centrally located, contracting from $\sim$14 Mpc to 2–3 Mpc over the same redshift range. HyLIRGs occupy the most compact regions of all: at $z\sim3.7$ they lie at distances of $\sim$6 Mpc, compared to $\sim$19 Mpc for ULIRGs at the same redshift, and at $z\sim2.2$ they are found at $\sim$13 Mpc versus $\sim$18 Mpc for ULIRGs. Together, these trends indicate that the most intense dusty star formation occurs preferentially in the densest regions during cluster growth. In projection, rich clusters reach projected angular radii of up to about 16 arcmin at $z\sim0$ and become progressively smaller toward higher redshift (9 arcmin at $z\sim5$), whereas poor proto-clusters remain much more compact, with typical projected angular radii of 4–5 arcmin and only weak redshift evolution. These characteristic spatial and angular scales provide a useful reference for designing future surveys and follow-up observations, helping to optimise search apertures and field-of-view choices for efficiently identifying proto-cluster environments and minimising projection effects from large-scale structure.

\item By tracking the three-dimensional distance of the ULIRGs at the 75th percentile (after ranking ULIRGs from faintest to brightest) from the proto-cluster centre, we find that at $z>1.5$ these galaxies are typically located within the inner 10–65 per cent of the cluster radius in rich clusters, and often within the inner 20-50 per cent in poor clusters. At $z<1.5$, as ULIRGs become rare and clusters approach dynamical relaxation, this central concentration disappears, and their spatial distribution becomes similar to that of the full galaxy population.

\item Star-forming galaxy pairs are abundant in the mock (42.5 per cent out of 585,334 associated pairs). However, only a small fraction of these pairs are above the TolTEC LSS survey detection thresholds: 0.42 per cent, 0.11 per cent, and 0.01 per cent of all pairs at 1.1, 1.4, and 2.0 mm, respectively. Among these detectable systems, only 36 per cent, 16 per cent, and $\sim$10 per cent are resolved given the TolTEC angular resolution, indicating that most detectable pairs will appear as blended sources, particularly at longer wavelengths, leading to an overestimation of their flux densities.

\item The LSS TolTEC survey will detect preferentially infrared-luminous galaxies ($L_{\rm IR}>10^{12}\; {\rm L}_{\sun}$). At 1.1 mm, about 86 per cent of detected sources will be ULIRGs and 3 per cent HyLIRGs; at 1.4 mm these fractions rise to 87 and 6 per cent; and at 2.0 mm to 82 and 14 per cent, respectively, with median redshifts of $z\simeq2.9$, 3.1 and 3.3. Several hundred sources will be strongly lensed ($\mu>10$).

\item Over a sky area of 60 deg$^2$, the expected area of the LSS TolTEC survey, we identify 72 proto-clusters associated with rich clusters (48 at $z \geq 2$ and 19 at $z \geq 4$), considering the full catalogue without applying any flux or resolution cuts. When the TolTEC detection limits are applied, the number of detectable galaxies per proto-cluster increases with redshift. At 1.1 mm, the median number rises from $\sim 3$ galaxies at $z \sim 2.2$ to a peak of $\sim 8$ at $z \sim 3.2$, and then declines to $\sim 2$ by $z \sim 5.5$.

\item Through an analysis of the angular separations between proto-cluster members detectable by TolTEC, we find that typical separations are large, with median values of $\sim5$--$6.6$ arcmin at $z\sim2$--3.5, decreasing to $\sim3.8$ arcmin at higher redshift. These scales are significantly larger than the TolTEC beam size (5 arcsec), indicating that the TolTEC LSS survey has sufficient angular resolution to spatially resolve the vast majority of detectable proto-cluster members. However, its sensitivity remains the main limitation, with only a small fraction of the proto-cluster galaxy population expected to be detected. Deeper observations will therefore be required to recover a larger fraction of the dusty star-forming galaxy population in proto-clusters.
\end{enumerate}

\section*{Acknowledgements}

This research was supported by a Mexican PhD scholarship from the Secretaría de Ciencia, Humanidades, Tecnología e Innovación (SECIHTI). This work was also funded by SECIHTI through project A1-S-45680.  IA is supported by Grant ATR2024-154316 funded by MICIU/AEI/10.13039/501100011033 and PID2022-136598NB-C33 by MCIN/AEI/10.13039/501100011033 and by “ERDF A way of making Europe”.

\section*{Data Availability}

The mock redshift survey is publicly available through \href{https://mnemosyne.inaoep.mx/index.php/s/phXy5KLx09FumYw} {https://mnemosyne.inaoep.mx/index.php/s/phXy5KLx09FumYw}. This release includes the GARDENS-Wide (100 square degrees) catalogue together with an updated version of the GARDENS-Deep (5.3 square degrees) catalogue. The catalogues are also available upon request from the corresponding author.



\bibliographystyle{mnras}
\bibliography{bibliography} 





\bsp	
\label{lastpage}
\end{document}